\documentclass{emulateapj}

\usepackage{graphicx}
\usepackage{psfig}
\usepackage{apjfonts}

\def\cm{\textrm{cm}}
\def\mm{\textrm{mm}}
\def\micron{\mu\textrm{m}}

\def\erg{\textrm{erg}}
\def\kpc{\textrm{kpc}}
\def\pc{\textrm{pc}}
\def\Mpc{\textrm{Mpc}}

\def\Kelv{\textrm{K}}

\def\ergps{\textrm{erg}~\textrm{s}^{-1}}
\def\gcm2{\textrm{g}~\textrm{cm}^{-2}}

\def\kms{\textrm{km}~\textrm{s}^{-1}}

\def\phcm2s1{\textrm{photons}~\textrm{cm}^{-2}~\textrm{s}^{-1}}

\def\eV{\textrm{eV}}
\def\keV{\textrm{keV}}
\def\MeV{\textrm{MeV}}
\def\GeV{\textrm{GeV}}

\def\yr{\textrm{yr}}
\def\Myr{\textrm{Myr}}

\def\Msun{\textrm{M}_{\sun}}
\def\Lsun{\textrm{L}_{\sun}}
\newcommand{\mean}[1]{\ensuremath{\langle #1 \rangle}}

\begin{document}

\title{Gamma-Ray Dominated Regions: Extending the Reach of Cosmic Ray Ionization in Starburst Environments}
\author{Brian C. Lacki\altaffilmark{1,2}}
\altaffiltext{1}{Jansky Fellow of the National Radio Astronomy Observatory}
\altaffiltext{2}{Institute for Advanced Study, Einstein Drive, Princeton, NJ 08540, USA}

\begin{abstract}
Cosmic rays are appealing as a source of ionization in starburst galaxies because of the great columns they can penetrate, but in the densest regions of starbursts, they may be stopped by pion production and ionization energy losses.  I argue that gamma rays are the source of ionization in the deepest molecular clouds of dense starbursts, creating Gamma-Ray Dominated Regions (GRDRs).  Gamma rays are not deflected by magnetic fields, have a luminosity up to $\sim 1/3$ that of the injected cosmic rays, and can easily penetrate column depths of $\sim 100\ \gcm2$ before being attenuated by $\gamma Z$ pair production.  The ionization rates of GRDRs, $\la 10^{-16}\ \sec^{-1}$, are much smaller than in cosmic ray dominated regions, but in the most extreme starbursts, they may still reach values comparable to those in Milky Way molecular clouds.  The gas temperatures in GRDRs could be likewise low, $\la 10\ \Kelv$ if there is no additional heating from dust or turbulence, while at high densities, the kinetic temperature will approach the dust temperature.  The ratio of ambipolar diffusion time to free-fall time inside GRDRs in dense starbursts is expected to be similar to those in Milky Way cores, suggesting star-formation can proceed normally in them.  The high columns of GRDRs may be opaque even to millimeter wavelengths, complicating direct studies of them, but I argue that they could appear as molecular line shadows in nearby starbursts with ALMA.  Since GRDRs are cold, their Jeans masses are not large, so that star-formation in GRDRs may have a normal or even bottom-heavy initial mass function.
\end{abstract}

\keywords{galaxies:starburst -- galaxies:ISM -- gamma rays: galaxies -- cosmic rays -- galaxies: star formation}

\section{Introduction}
\label{sec:Introduction}

With large star-formation rates often concentrated into small volumes, starburst galaxies are filled with the many kinds of high energy radiation that accompany star-formation, from ultraviolet photons to cosmic rays (CRs).  This radiation can interact with the interstellar medium (ISM) of the abundant gas to heat or ionize it.  The ionization of gas in star-forming environments in particular can drive chemistry in molecular clouds and sets the dynamics of magnetic fields that may in turn regulate star formation.  

However, the ionization rate from this radiation is limited by two key factors: the luminosity of the radiation, and the regions within the starburst it can penetrate into.  Energy conservation ultimately limits the rate atoms can be ionized by radiation; more radiation generated means more gas can be ionized, subject to the efficiency with which that radiation is converted into ionization.  A powerful source of ionizing luminosity are extreme ultraviolet (EUV; $h\nu \ge 13.6\ \eV$) or Lyman continuum photons from young O stars.  The O stars create H II regions around themselves, where the gas is highly ionized.  Indeed, the ionizing photon rate as measured by free-free radio luminosity is correlated with the star-formation rate \citep[e.g.,][]{Condon92,Murphy11}.  The amount of ionized material in H II regions is large because of the high efficiency that EUV light ionizes surrounding gas: starburst galaxies are highly optically thick to EUV light \citep[e.g.,][]{Leitherer95,Hurwitz97,Heckman01}.  Almost all of the power in EUV radiation goes into ionizing H II regions or possibly into dust absorption.

However, the high optical depth of neutral gas to EUV light also poses a problem for the ionization of most of the ISM.  Most of the EUV light is downgraded before it goes very far, so that H II regions are a small fraction of the volume of starburst galaxies, and something else must be responsible for ionizing the dense molecular gas in starburst galaxies.  Far ultraviolet (FUV; $6\ \eV \le h\nu \le 13.6\ \eV$) light is not absorbed by neutral hydrogen, but it can dissociate molecular hydrogen and ionize carbon: thus, regions where FUV light penetrates into are photodissociation regions (PDRs; see the extensive review by \citealt{Hollenbach99}).  PDRs include much of the neutral gas in the Milky Way, but FUV light is quickly absorbed by dust (limiting both the power going into ionization and the penetration depth), and PDRs extend only to columns of $N_H \approx 2 \times 10^{22}\ \cm^{-2}$ ($0.03\ \gcm2$), smaller than typical columns through starbursts ($0.1 - 10\ \gcm2$; \citealt{Kennicutt98,Hopkins10}).

The interiors of molecular clouds are ionized by more penetrative but less luminous sources of radiation.  X-rays can traverse higher columns and can ionize neutral gas, creating X-ray Dominated Regions (XDRs; \citealt{Maloney96,Meijerink05}).  In active galactic nuclei with high X-ray luminosities, X-rays can provide most of the ionizing luminosity, although this is concentrated towards the nucleus \citep{Papadopoulos10-CRDRs}.  Even in pure starburst galaxies, hard X-rays are emitted by high mass X-ray binaries, with a typical luminosity of $L_X \approx 10^{-4} L_{\star}$, where $L_{\star}$ is the bolometric (largely in the infrared) luminosity from star-formation \citep[e.g.,][]{Ranalli03,Persic04}.  However, most galaxies are optically thin to hard X-rays, meaning most of that luminosity simply leaves the starburst before ionizing neutral gas.  Furthermore, even hard X-rays are stopped once the columns become Compton thick ($N_H \approx 10^{24} \cm^{-2}$; $2.5\ \gcm2$), which can occur in extreme starbursts like Arp 220 \citep{Lehmer10} and in molecular clouds.  Even higher column densities can be attained in the parsec-scale gas around active galactic nuclei \citep[][and references therein]{Hopkins12}.

An attractive candidate for the ionization of starbursts is cosmic rays \footnote{In general, when I refer to CRs in this paper, I mean CR protons (and other nuclei) \citep{Suchkov93,Papadopoulos10-CRDRs}, which make up the bulk of the CR power in the Milky Way.  However, low energy CR electrons and positrons can also contribute to the CR ionization rate.}, which ionize the interiors of Milky Way molecular clouds.  The CRs regulate the temperature and ionization fraction in molecular clouds and cores, and because they pervade Milky Way molecular clouds, they set uniform conditions for star-formation throughout the Galaxy.  CRs have many advantages as ionization agents in starbursts.  (1) They are produced by star-formation at similar luminosities as X-rays, with $L_{\rm CR} \approx 3 \times 10^{-4} L_{\star}$ \citep{Thompson07}.  Since they trace star-formation, they naturally should be present in star-forming regions, unlike X-rays from AGNs, which are intense only near the nucleus itself.  (2) They can traverse columns of up to $\sim 80\ \gcm2$ at GeV energies, stopped mainly by inelastic collisions with ISM atoms that produce pions.  At lower kinetic energies (less than a few hundred MeV), ionization losses also stop them, becoming rapidly more effective at smaller kinetic energies.  They can therefore penetrate into regions of high dust extinctions, a big advantage over UV photons in the dusty environments of starbursts.  (3) Galaxies are extreme scattering atmospheres for CRs, deflecting them with magnetic inhomogeneities.  This allows the CR energy density to build up to very high levels \citep[e.g.,][]{Socrates08}.  Thus, \citet{Papadopoulos10-CRDRs} proposed that molecular gas in starburst galaxies are Cosmic Ray Dominated Regions (CRDRs; see also \citealt{Suchkov93}).  

But the high scattering optical depth to CRs is a double-edged sword.  On the one hand, it means that each CR has many chances to ionize gas atoms as it scatters across a starburst, but on the other it means that a CR has many chances to be destroyed through pionic and ionization losses.  The scattering atmosphere means that CRs see a much higher column that can stop them than photons do.  Indeed, in the Milky Way, there is evidence that low energy ($<100\ \MeV$) CR components contribute to high ionization rates in the diffuse ISM but are stopped by their own ionization losses \citep{Indriolo09}.  In dense starburst galaxies, protons may even lose all of their energy to pionic (and ionization) losses, in the ``proton calorimeter'' limit -- with little of the CR luminosity escaping \citep{Pohl94,Loeb06,Thompson07,Lacki10-FRC1}.  So while starbursts with their dense gas may efficiently convert CR energy into ionization, CRs may be unable to reach heavily shielded molecular regions in starbursts.  The GeV and TeV gamma-ray observations of M82 and NGC 253 \citep{Acero09,Acciari09,Abdo10-Starburst} indicate that roughly $f_{\pi} \sim 20 - 40\%$ of their CR power is lost to pionic emission \citep{Lacki11-Obs}, much higher than in the Milky Way where the ratio is only a few percent \citep[e.g.,][]{Strong10}, implying that pionic losses may be important for CRs in starbursts \citep{Lacki10-FRC1}.  

An even more worrying problem with CR ionization in starbursts are the presence of winds \citep[e.g.,][]{Chevalier85,Heckman03}: these carry CRs out of their starbursts \citep[c.f.,][]{Suchkov93}, effectively wasting the power they could provide in ionizing the ISM; but unlike diffusion, winds do not let them penetrate deeper into molecular clouds \citep{Crocker11-Wild}.  The hard gamma-ray spectra of M82 and NGC 253 \citep{Acero09,Acciari09,Abdo10-Starburst}, combined with the evidence that $f_{\pi} \la 1$, suggest that winds are important for CR transport in them, though denser starbursts like Arp 220 may be true proton calorimeters.  However, \citealt{Suchkov93} and \citealt{Papadopoulos10-CRDRs} did assume that winds limited the lifetimes of CRs in starbursts when calculating the CR energy densities.

\begin{figure}
\centerline{\includegraphics[width=8cm]{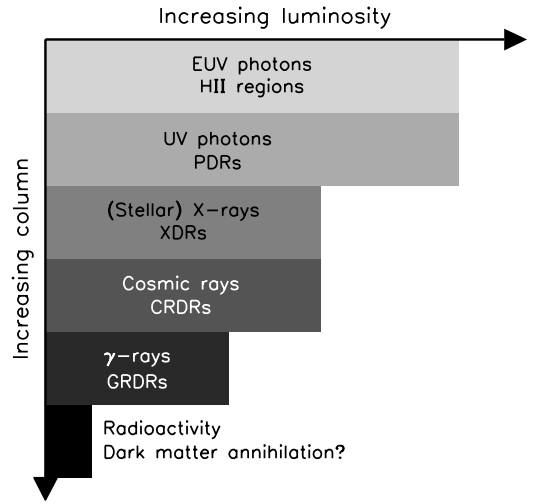}}
\figcaption[figure]{Schematic diagram showing the kinds of ionizing radiation in star-forming galaxies, and the regions they can ionize.  UV light has a great luminosity, but cannot penetrate far into the starburst; gamma rays are low luminosity but can penetrate deep into molecular clouds.  Cosmic rays and X-rays from X-ray binaries are intermediate in both luminosity and penetration depth.  The figure is not to scale, and the units are arbitrary.\label{fig:Explanation}}
\end{figure}

To summarize the problem for ionizing radiation, radiation is optimally effective in uniformly ionizing regions when $\tau = 1$: in media with smaller $\tau$, the radiation escapes without efficiently ionizing the gas, but in media with larger $\tau$, the radiation ionizes only the skin of the region and leaves a shielded core.  We therefore expect a hierarchy of ionization regions in starburst galaxies, with each step having lower luminosity but higher penetration depth.  The H II regions and PDRs from UV light receive the most ionizing power but only constitute a small volume of starbursts; XDRs and CRDRs receive much less power but pervade denser phases of starbursts (Figure~\ref{fig:Explanation}).  Is there anything beyond CRDRs, which could guarantee the ionization of the molecular cores of starbursts?  

I argue here that this ionization could come from gamma rays.  High energy gamma rays are produced in starbursts by the pionic losses of protons.  Additional leptonic gamma rays come from electrons and positrons ($e^{\pm}$), some of which may be secondary $e^{\pm}$ made by pion production.  Unlike CRs, gamma rays are not deflected by magnetic fields, and they are stopped only by Bethe-Heitler $\gamma Z$ pair production at columns of $\sim 10^{26}\ \cm^{-2}$ ($\sim 160\ \gcm2$).  In $\gamma Z$ pair production, the gamma ray produces an electron and a positron in the electric field of an atomic nucleus ($Z + \gamma \to Z + e^- + e^+$) \citep[e.g.,][]{Berestetskii79}.  The result in dense columns is a cascade: the $e^{\pm}$ either ionize the ISM directly if they are low enough energy, or effectively downgrade the energy of the initial gamma ray when they cool by radiating.  Thus it is possible to use the proton calorimetry of a starburst against it: the gamma rays can act as a ``second wind'' for CR ionization \citep[c.f.,][]{Umebayashi81}.  The result is that even if CRs are stopped in the outer reaches of a molecular cloud, the inner regions can still be ionized as a Gamma-Ray Dominated Region (GRDR; Figure~\ref{fig:Explanation}).  Of course, in most of the starburst, gamma rays escape most of the time, with most of the energy ``wasted'' that way, but the key point is that gamma rays provide a low level of ionization that extends extremely deep into the molecular clouds of a starburst.

This paper will consider basic properties of GRDRs in extreme starburst environments.  Using the energetics of gamma-ray production in starbursts, I estimates the characteristics of GRDRs in section~\ref{sec:BasicProperties}, including the ionization rate (section~\ref{sec:Estimates}), ionization fraction and ambipolar diffusion time (section~\ref{sec:GRDRxe}), and temperature (section~\ref{sec:GRDRTemperature}).  I then estimate when the transition from CR to gamma-ray ionization occurs and demonstrate how the extreme scattering atmospheres of starburst can turn even clouds with relatively small columns into GRDRs (section~\ref{sec:ScatteringEffect}).  To more accurately calculate how the pair $e^{\pm}$ ionize the ISM, I construct a simple 1D model of a $\gamma Z$ cascade in a starburst in section~\ref{sec:CascadeModel}.  I consider how GRDRs may appear observationally, including their indirect effects on stellar initial mass functions in section~\ref{sec:Observations}.  I finally discuss some additional speculations in the Conclusions, section~\ref{sec:Conclusions}.

\section{Basic Characteristics of GRDR\lowercase{s}}
\label{sec:BasicProperties}
\subsection{Estimates of GRDR Ionization Rates}
\label{sec:Estimates}

Suppose a starburst is a disk with radius $R$ and a midplane-to-edge scale height $h$.  It has a gamma ray luminosity of $L_{\gamma}$, giving a gamma-ray energy flux of $F_{\gamma} = L_{\gamma} / (2 \pi R^2)$.  The volumetric ionization power that gamma rays deposit in a gas cloud that is optically thin to gamma rays ($\tau_{\gamma Z} < 1$) is
\begin{equation}
\label{eqn:HeatingRate}
\Gamma_{\rm ion}^{\gamma} = F_{\gamma} \tau_{\gamma Z} f_{\rm ion}^{\gamma} / \ell = F_{\gamma} N_H \sigma_{\gamma Z} f_{\rm ion}^{\gamma} / \ell = F_{\gamma} n_H \sigma_{\gamma Z} f_{\rm ion}^{\gamma}
\end{equation}
where $\ell$ is the gas column length, $n_H$ is the number density of hydrogen atoms\footnote{Throughout this paper, I use the number density of hydrogen atoms, $n_H$, not the number density of hydrogen molecules $n(H_2)$.  The conversion between the two is $n_H = 2 n(H_2)$.  Likewise, the column density $N_H$ is for hydrogen atoms, not hydrogen molecules.}, $N_H = n_H \ell$ is the column density, $\sigma_{\gamma Z}$ is the cross section for $\gamma Z$ pair production, and $f_{\rm ion}^{\gamma}$ is the fraction of energy in pair $e^{\pm}$ that goes into ionization losses.  The $f_{\rm ion}^{\gamma}$ quantity depends on the physical conditions within the cloud and the spectrum of the gamma rays; here I will estimate it as $f_{\rm ion}^{\gamma} = 0.1$.

When a typical high energy proton ionizes an atom, the ejected electron will have a typical kinetic energy of $\sim 35\ \eV$ \citep[e.g.,][]{Cravens78}; the CR proton therefore typically loses $(35 + 13.6) = 49\ \eV$ per ionization event.  On the other hand, the initial ionized electron in turn ionizes, on average, $\phi = 0.7$ additional electrons \citep[e.g.,][]{Cravens78}.  Therefore, if ionizations by high energy $e^{\pm}$ are similar to those by high energy protons, each $e^{\pm}$ requires $E_{\rm ion} = 49\ \eV / (1 + \phi) = 29\ \eV$ per ionization event, whether primary or secondary.  Here I scale $E_{\rm ion}$ to 30 eV for simplicity.

The ionization rate from gamma rays is then simply $\zeta_{\gamma} \approx \Gamma_{\rm ion}^{\gamma} / (n_H E_{\rm ion})$:
\begin{equation}
\label{eqn:zetaGamma}
\zeta_{\gamma} \approx \frac{F_{\gamma} \sigma_{\gamma Z} f_{\rm ion}^{\gamma}}{E_{\rm ion}}.
\end{equation}

Few starbursts are gamma-ray detected, but we can estimate the gamma-ray flux using the star-formation rate, the Schmidt law, and the time scale for pionic energy losses.  The gamma-ray luminosity is
\begin{equation}
L_{\gamma} \approx f_{\pi} L_{\rm CR} / 3,
\end{equation}
where $f_{\pi}$ is the fraction of the CR power that goes into pionic losses: $f_{\pi} \approx t_{\rm life} / t_{\pi} = [1 + t_{\pi}/t_{\rm wind}]^{-1}$ at high energies.  The pionic loss time is roughly 
\begin{equation}
\label{eqn:tPi}
t_{\pi} \approx 50\ \Myr \left(\frac{\mean{n_H}}{\cm^{-3}}\right)^{-1},
\end{equation}
where $\mean{n_H}$ is the mean density of gas in the entire starburst \citep{Mannheim94}; this is equivalent to saying pionic losses stop CR protons over a column $\Sigma_{\pi} = t_{\pi} n_H c m_H = 79\ \gcm2$.  In terms of the mean column density of gas in the starburst $\mean{\Sigma_g} = 2 m_H h n_H$, we have $t_{\pi} \approx 100\ \Myr\ m_H h / \mean{\Sigma_g}$.  The wind escape time is $t_{\rm wind} = h / v_{\rm wind}$.  Combining these timescales, I have:
\begin{equation}
\label{eqn:FCal}
f_{\pi} \approx \left[1 + 0.16 \left(\frac{v_{\rm wind}}{300\ \kms}\right)\left(\frac{\mean{\Sigma_g}}{\gcm2}\right)^{-1}\right]^{-1}.
\end{equation}

The luminosity of injected CRs scales with the supernova rate (or more generally, the massive star formation rate).  A CR luminosity of $10^{50}\ \erg$ per supernova is expected.  For a Salpeter IMF, we expect a supernova rate of $\Gamma_{\rm SN} = 0.0064\ \yr^{-1} ({\rm SFR}/\Msun\ \yr^{-1})$ \citep{Thompson07}.  In turn, I can relate the star-formation rate to the gas density using the Schmidt law, $\Sigma_{\rm SFR} = {\rm SFR} / (\pi R^2) = 2.5 \times 10^{-4} \Msun\ \yr^{-1}\ \kpc^{-2} (\mean{\Sigma_g}/\Msun\ \pc^{-2})$ \citep{Kennicutt98}.  Putting these relations all together, I find the gamma-ray flux:
\begin{eqnarray}
F_{\gamma} & \approx & 0.0036\ \erg \sec^{-1} \cm^{-2}\ f_{\pi} \left(\frac{\rm SFR}{\Msun\ \yr^{-1}}\right) \left(\frac{R}{250\ \pc}\right)^{-2}\\
F_{\gamma} & \approx & 0.0252\ \erg \sec^{-1} \cm^{-2}\ f_{\pi} \left(\frac{\mean{\Sigma_g}}{\gcm2}\right)^{1.4}
\end{eqnarray}
and finally,
\begin{equation}
F_{\gamma}  \approx \left\{ \begin{array}{ll} 0.157\ \erg \sec^{-1} \cm^{-2} \left(\frac{\mean{\Sigma_g}}{\gcm2}\right)^{2.4} & (\mean{\Sigma_g} \la 0.16\ \gcm2)\\
0.0252\ \erg \sec^{-1} \cm^{-2} \left(\frac{\mean{\Sigma_g}}{\gcm2}\right)^{1.4} & (\mean{\Sigma_g} \ga 0.16\ \gcm2) \end{array} \right.
\end{equation}

\begin{figure}
\centerline{\includegraphics[width=8cm]{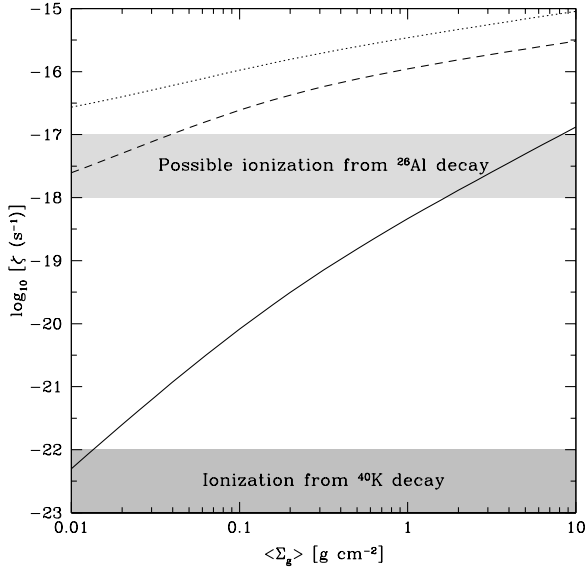}}
\figcaption[figure]{Plot of ionization rate from gamma rays (solid) and CRs (dotted: momentum power law injection; dashed: total energy power law injection), using the Schmidt Law.  The GRDR ionization rate is much smaller than the CRDR ionization rate at all surface densities, but the ratio of $\zeta_{\gamma} / \zeta_{\rm CR}$ grows with $\mean{\Sigma_g}$.  Gamma-ray ionization is more important than $^{40}$K radioactivity in virtually all starbursts, and approaches the $\zeta$ value in Milky Way molecular clouds at the highest gas surface densities.  Individual galaxies may have higher or lower $\zeta$ depending on their deviation from the Schmidt Law.  I have assumed $v_{\rm wind} = 300\ \kms$  and $f_{\rm ion}^{\gamma} = 0.1$.\label{fig:GRDRZeta}}
\end{figure}

Evaluating the cross section for $\gamma Z$ pair production \citep{Berestetskii79},
\begin{equation}
\sigma_{\gamma Z} = \frac{28}{9} \alpha_{\rm FS} r_e^2 \left(\ln \frac{2E_{\gamma}}{m_e c^2} - \frac{109}{42}\right)
\end{equation}
at $E_{\gamma} = 1\ \GeV$ to be $1.02 \times 10^{-26}\ \cm^2$, I find
\begin{equation}
\zeta_{\gamma} \approx \left\{ \begin{array}{r} \displaystyle 3.3 \times 10^{-18} \sec^{-1} \left(\frac{\mean{\Sigma_g}}{\gcm2}\right)^{2.4} \left(\frac{f_{\rm ion}^{\gamma}}{0.1}\right) \left(\frac{E_{\rm ion}}{30\ \eV}\right)^{-1}\\
(\mean{\Sigma_g} \la 0.16\ \gcm2)\\\\
\displaystyle 5.3 \times 10^{-19} \sec^{-1} \left(\frac{\mean{\Sigma_g}}{\gcm2}\right)^{1.4} \left(\frac{f_{\rm ion}^{\gamma}}{0.1}\right) \left(\frac{E_{\rm ion}}{30\ \eV}\right)^{-1} \\
(\mean{\Sigma_g} \ga 0.16\ \gcm2)\end{array} \right.
\end{equation}
The results are plotted in Figure~\ref{fig:GRDRZeta}.  These ionization rates are small, though they do rise to $\sim 10^{-17}\ \sec^{-1}$ for the densest ULIRGs with $\mean{\Sigma_g} \approx 10\ \gcm2$.  In principle, if $f_{\rm ion}$ is 1, these ionization rates can be ten times higher.  In any case, though, they remain much higher than the ionization rate from long-lived radioactive elements such as $^{40}$K in the Milky Way, $\zeta_R \approx 10^{-22}\ \sec^{-1}$ \citep{Umebayashi81,Umebayashi09} even for mean column densities of $\mean{\Sigma_g} \approx 0.01\ \gcm2$.  

In starbursts, there may be additional radioactive ionization from relatively short-lived isotopes like $^{26}$Al because of the high supernova rate \citep[c.f.,][]{Diehl06}; for example, in the young Solar System, $^{26}$Al alone sustained an ionization rate of $\sim 10^{-18}\ \sec^{-1}$ (\citealt{Umebayashi09}; see also \citealt{Stepinski92}).  In a companion paper \citep{Lacki12-Al26}, I find that the mean $^{26}$Al ionization rate in starbursts quite high because of their rapid specific star-formation rates, of the order $10^{-18} - 10^{-17}\ \sec^{-1}$.  If true, gamma-ray ionization only dominates in the most extreme starbursts with $\mean{\Sigma_g} \approx 10\ \gcm2$.  However, whether these high ionization rates are actually attained in the molecular gas of starbursts depends on whether the $^{26}$Al is actually well-mixed with the gas; as with CRs, the propagation of $^{26}$Al from its sources matters.  This is not an issue with the gamma rays.  

Some starbursts are in fact gamma-ray detected, allowing us to directly calculate $F_{\gamma}$.  The brightest starburst in the gamma-ray sky in terms of observed flux is M82, with a GeV-TeV gamma-ray luminosity of $2 \times 10^{40}\ \ergps$ \citep{Abdo10-Starburst} and a starburst radius of 250 pc \citep[e.g.,][]{Goetz90,Williams10}.  Therefore, using eqn.~\ref{eqn:zetaGamma}, I find 
\begin{equation}
\zeta_{\gamma} = 1.1 \times 10^{-19}\ \sec^{-1} \left(\frac{f_{\rm ion}}{0.1}\right) \left(\frac{E_{\rm ion}}{30\ \eV}\right)^{-1},
\end{equation}
much smaller than even in Milky Way molecular clouds, but still much larger than the ionization rate from long-lived radioisotopes like $^{40}$K.  

Arp 220, which if star-formation powered has a star-formation rate of $200 - 300\ \Msun\ \yr^{-1}$ (\citealt{Torres04} and references therein; \citealt{Anantharamaiah00}), is a more likely home for GRDRs.  Most of its radio emission comes from two 50 - 100 pc radius nuclei \citep{Norris88,Downes98}, although there are varying estimates of the literature of how much of the infrared luminosity is actually from the nuclei \citep{Downes98,Downes07,Sakamoto08}.  If I assume that each nucleus has a star-formation rate of $50\ \Msun\ \yr^{-1}$ (for a bolometric luminosity of $2.8 \times 10^{11}\ \Lsun$ each) and that Arp 220 is a proton calorimeter, then I find:
\begin{equation}
\zeta_{\gamma} \approx 4.8 \times 10^{-17}\ \sec^{-1} \left(\frac{R}{50\ \pc}\right)^{-2} \left(\frac{f_{\rm ion}}{0.1}\right) \left(\frac{E_{\rm ion}}{30\ \eV}\right)^{-1}.
\end{equation}
Thus, even deep molecular cores in Arp 220's nuclei should have an ionization rate comparable to the molecular clouds of the Milky Way.

A further effect which may enhance the GRDR ionization rate is clumpiness in the gamma-ray emission of the starburst.  I assumed that the gamma-ray emission was evenly distributed across the starburst.  However, some molecular clouds in starbursts may be located near gamma-ray sources.  The outer layers of the molecular cloud likely themselves generate gamma rays as CRs are stopped by them, enhancing the interior gamma-ray flux and gamma-ray ionization rate \citep[e.g.,][]{Umebayashi81}.

\subsection{Comparison with CRDR Ionization Rates}
\label{sec:CRDRComparison}
How does the ionization rate in GRDRs compare with that in the surrounding CRDRs?  One way of estimating the ionization rate from CRs is to scale the Milky Way molecular cloud ionization rate by the energy density of CRs.  A potential problem with this approach, however, is that the CR spectrum has a different shape in starbursts than in the Milky Way.  Whereas the Milky Way has a $E^{-2.7}$ gamma-ray spectrum because CR protons escape through diffusion \citep[e.g.,][]{Ginzburg76}, starbursts are observed to have hard $E^{-2.2}$ gamma-ray spectra from strong advection and/or pionic losses for protons \citep{Acero09,Acciari09,Abdo10-Starburst}.  Therefore, some of the larger CR energy density in starbursts is in very high energy particles which are not responsible for ionization.\footnote{The spectrum of CRs can be approximated by multiplying the injection spectrum times a lifetime at each energy, and from this we can get the energy density of particles at, say, 1 GeV ($E dU/dE = E^2 dN/dE$).  Then in principle the ratio of the starburst and Milky Way $E dU/dE$ at $\la 1\ \GeV$ can be used to estimate the ionization rate.  This is effectively done in \citet{Suchkov93} and \citet{Papadopoulos10-CRDRs}: the CR energy density is related to the star-formation rates (injection) divided by an effective escape velocity, which is related to the GeV diffusive escape time in the Milky Way and the (fastest observed) wind speed in starbursts.  In any case, however, the calculation does require more care than simply computing the integrated energy density.}  Furthermore, this approach requires a CR ionization rate for the Milky Way and a Milky Way CR energy density.

Another way to estimate the CR ionization rate is to use the injected power in CR protons that is lost to ionization:
\begin{equation}
L_{\rm ion} = \int \frac{dQ}{dE} K f_{\rm ion}^p (K) dE,
\end{equation}
where $f_{\rm ion}^p (K)$ is the fraction of energy of a CR with kinetic energy $K$ that is lost to ionization and $dQ/dE$ is the injection spectrum of CR protons.  The fraction of energy going into ionization is roughly:
\begin{equation}
f_{\rm ion}^p \approx \frac{t_{\rm life}}{t_{\rm ion}} \approx \left[1 + \frac{t_{\rm ion}^p}{t_{\pi}} + \frac{t_{\rm ion}^p}{t_{\rm wind}}\right]^{-1}.
\end{equation}
Supposing the pionic losses to have a threshold of $K = 140\ \MeV$, the rest energy of a pion,
\begin{equation}
f_{\rm ion}^p \approx \left\{ \begin{array}{ll} \displaystyle \left[1 + \frac{t_{\rm ion}^p}{t_{\rm wind}}\right]^{-1} & (K < 140\ \MeV)\\ \displaystyle \left[1 + \frac{t_{\rm ion}^p}{t_{\pi}} \frac{1}{f_{\pi}} \right]^{-1} & (K > 140\ \MeV). \end{array} \right.
\end{equation}
The lifetime of protons to ionization losses is approximately
\begin{equation}
t_{\rm ion}^p \approx 159\ \Myr \left(\frac{n_H}{\cm^{-3}}\right)^{-1} \left(\frac{K}{\GeV}\right) \beta
\end{equation}
where $\beta = \sqrt{1 - 1/\gamma^2}$ is the speed of the proton divided by $c$ \citep[][and references therein]{Torres04}.  This gives us, roughly, an ionization to pionic lifetime ratio of $t_{\rm ion}^p / t_{\pi} \approx 3.18 (K / \GeV) \beta$.

Now suppose the injection spectrum is a power-law in total energy, extending from $E = m_p c^2$ ($K = 0$) to infinity: $dQ/dE = Q_0 (E / m_p c^2)^{-p}$.  The normalization of this power law is set by $L_{\rm CR} = \int_{mc^2}^{\infty} Q_0 (E - m_p c^2) (E / m_p c^2)^{-p} dE$, giving $Q_0 = L_{\rm CR} [1/(p-2) - 1/(p-1)]^{-1} (m_p c^2)^{-2}$. When $p = 2.2$ and $f_{\pi} = 1$, I find that $L_{\rm ion} = 0.047 L_{\rm CR}$.  Another commonly assumed injection spectrum is a power law in momentum.  For $p = 2.2$ and $f_{\pi} = 1$, with minimum kinetic energies ranging from 1 MeV to 100 MeV, I find $L_{\rm ion} = (0.10 - 0.15) L_{\rm CR}$.  I therefore adopt $L_{\rm ion} \approx 0.1 L_{\rm CR}$ as a fiducial value in the calculations below.

The CR ionization rate is 
\begin{equation}
\label{eqn:CRDRrate}
\zeta_{\rm CR} = \frac{\eta_{\rm ion} L_{\rm CR}}{E_{\rm ion} V \mean{n_H}} = \frac{\eta_{\rm ion} L_{\rm CR} m_H}{E_{\rm ion} M_H}
\end{equation}
after defining $\eta_{\rm ion} \equiv L_{\rm ion} / L_{\rm CR}$.  In this equation, $M_H$ is the gas mass in the starburst.  Scaling to values typical in Arp 220's nuclei, I find
\begin{equation}
\label{eqn:zetaCRArp220}
\zeta_{\rm CR} = 4.4 \times 10^{-15}\ \sec^{-1} \left(\frac{\rm SFR}{50\ \Msun\ \yr^{-1}}\right) \left(\frac{M_H}{4 \times 10^8\ \Msun}\right)^{-1} \left(\frac{E_{\rm ion}}{30\ \eV}\right)^{-1}.
\end{equation}
This is $\sim 100$ times that of Milky Way molecular clouds or an Arp 220 GRDR. 

A convenient expression for the ratio of the ionization rates in a GRDR to a CRDR is the ratio of equations~\ref{eqn:zetaGamma} and \ref{eqn:CRDRrate}:
\begin{equation}
\frac{\zeta_{\gamma}}{\zeta_{\rm CR}} \approx \frac{f_{\pi} \sigma_{\gamma Z} f_{\rm ion}^{\gamma} h \mean{n_H}}{3 \eta_{\rm ion}} \approx \frac{f_{\pi} \sigma_{\gamma Z} f_{\rm ion}^{\gamma} \mean{\Sigma_g}}{6 \eta_{\rm ion} m_H}.
\end{equation}
where I have assumed a disk geometry for the volume ($V = 2 \pi R^2 h$).  In Arp 220, this ratio reaches
\begin{equation}
\label{eqn:ZetaRatio}
\frac{\zeta_{\gamma}}{\zeta_{\rm CR}} \approx 0.010 f_{\pi} \left(\frac{\mean{\Sigma_g}}{10\ \gcm2}\right) \left(\frac{f_{\rm ion}^{\gamma}}{0.1}\right) \left(\frac{\eta_{\rm ion}}{0.1}\right)^{-1}.
\end{equation}
Note this ratio increases even once proton calorimetry is attained.  This is because at high densities, the pionic lifetime continues to decrease as $n_H$, so that the energy density of CRs per supernova decreases; the energy density of gamma rays per supernova instead remains constant.  Only when $\tau_{\gamma Z} \ge 1$ for the entire starburst does the gamma-ray flux also start to diminish.

\subsection{The Ionization Fraction of GRDRs: Consequences for Ambipolar Diffusion}
\label{sec:GRDRxe}
\begin{figure}
\centerline{\includegraphics[width=8cm]{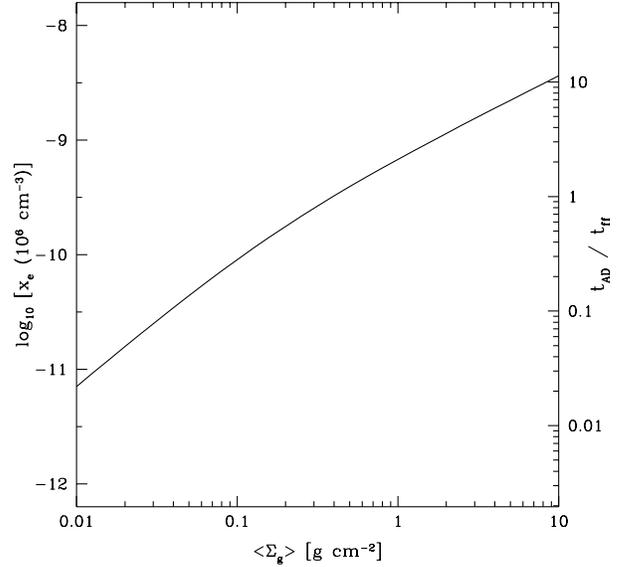}}
\figcaption[figure]{Plot of ionization fraction of $n_H = 10^6\ \cm^{-3}$ material in GRDRs, using the Schmidt Law.  The plot also shows the ratio of the ambipolar diffusion and free-fall times.  At high gas surface densities, ambipolar diffusion in GRDRs is slower than the free-fall time, suggesting that magnetic fields can regulate gas cloud collapse.  However, unlike in CRDRs, $t_{\rm AD} / t_{\rm ff} < 100$, so the collapse is not slowed far below observations.\label{fig:GRDRXe}}
\end{figure}

\citet{McKee89} gives the ionization fraction of a cloud with density $n_H$ as 
\begin{equation}
\label{eqn:xE}
x_e = 1 \times 10^{-7} r_{\rm gd}^{-1} \left(\frac{n_{\rm ch}}{n_H}\right)^{1/2} \left[\left(1 + \frac{n_{\rm ch}}{4 n_H}\right)^{1/2} + \left(\frac{n_{\rm ch}}{4 n_H}\right)^{1/2}\right].
\end{equation}
Here, $n_{\rm ch} \approx 1000 r_{\rm gd}^2 (\zeta / 10^{-17}\ \sec^{-1})$ is a characteristic density, and $r_{\rm gd}$ is the gas-to-dust ratio divided by 100 (see also \citealt{Papadopoulos10-CRDRs}).  For most GRDRs, the ionization rate is low enough that $n_{\rm ch} \ll n_H$ and this can be approximated as
\begin{equation}
x_e^{\rm GRDR} = 3.2 \times 10^{-9} \left(\frac{\zeta_{\gamma}}{10^{-17}\ \sec^{-1}}\right)^{1/2} \left(\frac{n_H}{10^6\ \cm^{-3}}\right)^{-1/2}
\end{equation}
The ionization fraction when $n_H = 10^6\ \cm^{-3}$ is plotted in Figure~\ref{fig:GRDRXe}.  In Arp 220, this fraction rises to $x_e^{\rm GRDR} \approx 10^{-8}$ for $n_H = 10^6\ \cm^{-3}$.  However, in using equation~\ref{eqn:xE}, I have ignored the possible appearence of different chemical effects in the very low ionization environments of GRDRs, such as charge being carried by dust grains \citep{McKee89}.

The ionization fraction is important for determining the ambipolar diffusion time, the time for the magnetic field to slip away from the neutral gas.  If the magnetic field remains strong in a cloud, free-falling gravitational collapse cannot proceed \citep[e.g.,][]{Mouschovias76}.  When the ambipolar diffusion time is longer than the free-fall time and the magnetic fields are strong, the collapse instead proceeds quasi-statically on the ambipolar diffusion time (e.g., \citealt{Mestel56}; see also the discussion in \citealt{Crutcher99}).  In most Milky Way starless cores, the ambipolar diffusion timescale is roughly an order of magnitude longer than the free-fall time \citep[e.g.,][]{Caselli98,Williams98,Maret07,Hezareh08}.  There is some recent evidence that at least some of the very deepest Milky Way cores may have low ionization fractions and ambipolar diffusion times approximately equal to the free-fall time \citep[e.g.,][]{Caselli02,Bergin07}.  Given that strong magnetic fields are observed by Zeeman splitting in dense OH maser regions in ULIRGs \citep{Robishaw08}, it seems plausible that magnetic fields regulate star formation in starbursts, so understanding the ambipolar diffusion rate is important.

\citet{McKee89} also gives the ambipolar diffusion time as $t_{\rm AD} \approx 1.6 \times 10^6 (x_e / 10^{-8})\ \yr$.  In the low ionization environments of GRDRs with $n_{\rm ch} \ll n_H$, we have
\begin{equation}
t_{\rm AD}^{\rm GRDR} = 5.1 \times 10^5\ \yr \left(\frac{\zeta_{\gamma}}{10^{-17}\ \sec^{-1}}\right)^{1/2} \left(\frac{n_H}{10^6\ \cm^{-3}}\right)^{-1/2}.
\end{equation}
We can compare this with the local free-fall time $t_{\rm ff} = \sqrt{3\pi / (32 G m_H n_H)}$:
\begin{equation}
t_{\rm ff} = 5.1 \times 10^4\ \yr \left(\frac{n_H}{10^6\ \cm^{-3}}\right)^{-1/2}.
\end{equation}
The interesting point for star-formation is that both the ambipolar diffusion time and free-fall time scale with $n_H^{-1/2}$.  Thus all GRDRs in a given starburst galaxy will have the same ratio of ambipolar and free-fall times, regardless of their overdensity.  This ratio is
\begin{equation}
\frac{t_{\rm AD}}{t_{\rm ff}} \approx 10 \left(\frac{\zeta_{\gamma}}{10^{-17}\ \sec^{-1}}\right)^{1/2}.
\end{equation}
The ratio is also plotted in Figure~\ref{fig:GRDRXe}; from the Schmidt Law, we expect ambipolar diffusion in GRDRs to be slower than free-fall in galaxies with $\mean{\Sigma_g} \ga 0.5\ \gcm2$.  Even in M82, this ratio is $\sim 1$, so that ambipolar diffusion is no quicker than free-fall time.  But in a starburst like Arp 220, this ratio reaches $22$.  

The ratio of ambipolar diffusion timescale to free-fall time in dense starburst GRDRs is therefore similar to those observed in Milky Way cores, $\sim 1 - 100$.  This means that star-formation may proceed similarly in dense starbursts and normal galaxies.  By contrast in CRDRs, the predicted ambipolar diffusion time is vastly longer than the free-fall time; star-formation would effectively be halted unless some mechanism, perhaps turbulence, speeds up ambipolar diffusion \citep{Papadopoulos10-CRDRs}.   It is interesting that in Arp 220, $t_{\rm AD}^{\rm GRDR}$ approaches $\sim 100 t_{\rm ff}$, the empirically observed timescale for star-formation \citep[e.g.,][]{Kennicutt98,Krumholz12}.  This suggests that in Arp 220, ambipolar diffusion in GRDRs may alter the star-formation efficiency slightly.  

\subsection{The Temperature of GRDRs}
\label{sec:GRDRTemperature}

\begin{figure}
\centerline{\includegraphics[width=8cm]{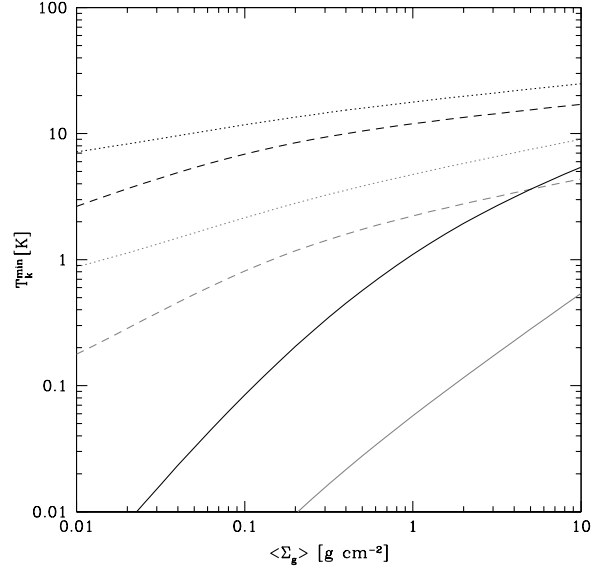}}
\figcaption[figure]{The minimum gas temperature in GRDRs and CRDRs using the Schmidt Law.   Black is for $n_H = 10^4\ \cm^{-3}$, and grey is for $n_H = 10^6\ \cm^{-3}$.  The solid lines are for GRDRs; dashed are for CRDRs with an energy power-law injection spectrum; dotted are for CRDRs with a momentum power-law injection spectrum.  Individual galaxies may be significantly off the plotted relations if they do not follow the Schmidt law.  \label{fig:GRDRTemp}}
\end{figure}

\citet{Papadopoulos10-CRDRs} gives a convenient expression for the minimum gas kinetic temperature of a CRDR which also applies for GRDRs, assuming no dust or turbulent heating:
\begin{equation}
\label{eqn:TkMinFull}
T_k^{\rm min} = 6.3\ \Kelv\ [(0.707 n_4^{1/2} \zeta_{-17} + 0.186^2 n_4^3)^{1/2} - 0.186 n_4^{3/2}]^{2/3}
\end{equation}
where $n_4 = n_H / (10^4\ \cm^{-3})$ and $\zeta_{-17} = \zeta / (10^{-17}\ \sec^{-1})$.  The temperatures for GRDRs and CRDRs are plotted in Figure~\ref{fig:GRDRTemp}.  

Simpler expressions can be found for $T_k^{\rm min}$ in the limit of high density and low ionization rate, where a Taylor series expansion yields:
\begin{equation}
T_k^{\rm min} = 9.7\ \Kelv\ n_4^{-2/3} \zeta_{-17}^{2/3}.
\end{equation}
Thus, using the expressions for $\zeta_{\gamma}$ above, I derive approximate gas kinetic temperatures:
\begin{equation}
T_k^{\rm min} \approx \left\{ \begin{array}{r} \displaystyle 0.22\ \Kelv \left(\frac{n_H}{10^6\ \cm^{-3}}\right)^{-2/3} \left(\frac{\mean{\Sigma_g}}{\gcm2}\right)^{1.6} \left(\frac{f_{\rm ion}^{\gamma}}{0.1}\right)^{2/3}\\
\displaystyle \times \left(\frac{E_{\rm ion}}{30\ \eV}\right)^{-2/3}\\
(\mean{\Sigma_g} \la 0.16\ \gcm2)\\\\
\displaystyle 0.064\ \Kelv \left(\frac{n_H}{10^6\ \cm^{-3}}\right)^{-2/3} \left(\frac{\mean{\Sigma_g}}{\gcm2}\right)^{0.93} \left(\frac{f_{\rm ion}^{\gamma}}{0.1}\right)^{2/3}\\
\displaystyle \times \left(\frac{E_{\rm ion}}{30\ \eV}\right)^{-2/3} \\
(\mean{\Sigma_g} \ga 0.16\ \gcm2) \end{array} \right.
\end{equation}

The minimum temperatures of GRDRs are clearly very low: only if $\mean{\Sigma_g}$ (and $\mean{\Sigma_{\rm SFR}}$) is very large and the local density is relatively small ($\la 10^6\ \cm^{-3}$) do the temperatures exceed 1 Kelvin.  For the values of $\zeta_{\gamma}$ derived for M82 in section~\ref{sec:Estimates}, I calculate $T_k^{\rm min}$ to be 0.5 K if $n_H = 10^4\ \cm^{-3}$ and 0.02 K if $n_H = 10^6\ \cm^{-3}$.  For Arp 220's nuclear starbursts, I find $T_k^{\rm min}$ is 9 K if $n_H = 10^4\ \cm^{-3}$ and 1.3 K if $n_H = 10^6\ \cm^{-3}$.  For comparison, the CRDR temperature in Arp 220 is $\sim 42\ \Kelv$ for $n_H = 10^4\ \cm^{-3}$ (using eqn.~\ref{eqn:zetaCRArp220} for $\zeta_{\rm CR}$).  It seems that in most starbursts GRDRs would be much colder than molecular clouds in the Milky Way.

Of course, the temperatures derived in equation~\ref{eqn:TkMinFull} are \emph{minimum} temperatures derived by assuming dust has a temperature of 0, so that the gas is cooled by collisions with dust grains.  In reality, clouds will be embedded in an intense far-infrared radiation field (and the CMB), heating its dust grains.  If the dust temperature is not cooled significantly by dust-grain interactions, if there is no source of dust heating -- whether light from young stars or absorption of molecular line emission -- within the GRDR, and if the dust is in thermal equilibrium with the external radiation field, there will be no net flux in dust thermal emission through the cloud because of energy conservation, so that the dust temperature should be equal to the external effective temperature.  

For a given dust temperature, dust-grain interactions will set a gas kinetic temperature floor where the dust-grain heating rate $\Gamma_{g-d}$ equals the molecular line cooling rate $\Lambda_{\rm line}$.  \citet{Papadopoulos10-CRDRs} quotes 
\begin{equation}
\Gamma_{g-d} = 2.5 \times 10^{-26} \left(\frac{n_H}{10^4\ \cm^{-3}}\right)^2 \left(\frac{T_k}{\Kelv}\right)^{1/2} \left(\frac{T_{\rm dust} - T_{k}}{\Kelv}\right)\ \erg\ \cm^{-2}\ \sec^{-1}
\end{equation}
and 
\begin{equation}
\Lambda_{\rm line} = 4.2 \times 10^{-24} \left(\frac{n_H}{10^4\ \cm^{-3}}\right)^{1/2} \left(\frac{T_k}{10\ \Kelv}\right)^3\ \erg\ \cm^{-2}\ \sec^{-1}.
\end{equation}
Equating these two rates, I find a temperature floor of
\begin{equation}
\label{eqn:TFloor}
T_{\rm floor} = 8.9\ \Kelv \left(\frac{n_H}{10^4\ \cm^{-3}}\right)^{3/5} \left(\frac{T_{\rm dust}}{40\ \Kelv}\right)^{2/5}
\end{equation}
when the density is low and $T_k \ll T_{\rm dust}$, and 
\begin{equation}
T_{\rm floor} = T_{\rm dust} - 1.7\ \Kelv \left(\frac{n_H}{10^6\ \cm^{-3}}\right)^{-3/2} \left(\frac{T_{\rm dust}}{40\ \Kelv}\right)^{5/2}
\end{equation}
when the density is high and $T_k \approx T_{\rm dust}$.  GRDRs with $n_H \ga 40000\ \cm^{-3}$ will have a kinetic temperature near the dust temperature when $T_{\rm dust} \approx 40\ \Kelv$.  At lower densities, if the heating and cooling rates are accurate, the gas will be much colder, though given the extreme conditions, the low temperatures should be approached with caution.  

Another possible source of heating is the dissipation of turbulence.  Most of the molecular gas in starbursts exists in a highly turbulent phase \citep[e.g.,][]{Downes98}.  Supersonic turbulence rapidly dissipates, partly because the supersonic motions generates shocks \citep[e.g.,][]{Stone98}.  The cores in molecular clouds and protostars achieve densities much higher than most of the mass in the turbulent ISM -- they are said to be ``decoupled'' from the surrounding medium -- although such collapsed structures can be initially generated by turbulence \citep[e.g.,][]{Klessen00}.  It is these clouds which have the high columns and densities where GRDRs are most likely to be present.  \citet{Bergin07} argues that in these clouds, turbulence is unimportant, based on the observed relationship between turbulent line width and cloud size in Milky Way molecular clouds \citep{Myers83,Larson05}. 

The cold temperatures of GRDRs have implications for their Jeans mass, which may set the minimum mass of stars in starbursts and affect their initial mass function (IMF).  The Jeans mass is
\begin{equation}
M_J = \left(\frac{k_B T_k}{G \mu m_H}\right)^{3/2} \left(n_H m_H\right)^{-1/2}.
\end{equation}
where $\mu = 0.75 \times 2 + 0.25 \times 4 = 2.5$ is the mean particle mass in the molecular ISM, in terms of the atomic mass of hydrogen $m_H$ \citep{Papadopoulos11-SF}.  In dust-heated gas with temperatures near $T_{\rm floor}$, I find the Jeans mass is
\begin{eqnarray}
\label{eqn:MJeansNumer}
M_J & \approx & 1.2\ \Msun \left(\frac{T_k}{9\ \Kelv}\right)^{3/2} \left(\frac{n_H}{10^4\ \cm^{-3}}\right)^{-1/2}\\
    & \approx & 1.1\ \Msun \left(\frac{T_k}{40\ \Kelv}\right)^{3/2} \left(\frac{n_H}{10^6\ \cm^{-3}}\right)^{-1/2}.
\end{eqnarray}
For a dust temperature of $40\ \Kelv$, the Jeans mass reaches a maximum of $1.8\ \Msun$ when $n_H \approx 1.3 \times 10^5\ \cm^{-3}$.  Thus, the Jeans mass of GRDRs is similar or perhaps slightly higher than that in normal galaxies, unlike in CRDRs, where it may be much higher ($\sim 10\ \Msun$) in starbursts \citep[c.f.][]{Papadopoulos11-SF}.

I conclude that the temperature of GRDRs is more likely set by the dust grains being illuminated by infrared radiation or possibly turbulence than gamma-ray heating, except perhaps in the lower density regions of the most extreme starbursts.

\section{The Transition from Cosmic Rays to Gamma Rays: The Perils of Cosmic Ray Scattering}
\label{sec:ScatteringEffect}
At first glance, the ionization rates and equilibrium temperatures of GRDRs seem surprisingly low compared to predictions for CRDRs, given the high gamma-ray luminosities of starbursts.  In a proton calorimetric starburst, $L_{\gamma}$ is about one-third of $L_{\rm CR}$, so why doesn't equation~\ref{eqn:HeatingRate} imply that the CRDR heating rate and ionization rate are also low?  The reason is that galaxies are extreme scattering atmospheres for CRs, at least for clouds of sufficiently low CR absorption optical depth.  The flux in equation~\ref{eqn:HeatingRate} is not the net flux, which is relatively small in galaxies, but the \emph{total} flux, which is much larger.  Because of the large scattering optical depth of a galaxy, a CR can cross a cloud multiple times: this increases the odds that it will ionize an atom in it, and therefore increases the CR ionization and heating rate.  However, the disadvantage of this is that a CR has a much greater chance of being destroyed by pion production or ionization cooling, precisely because it can cross the clouds multiple times.

\begin{figure}
\centerline{\includegraphics[width=8cm]{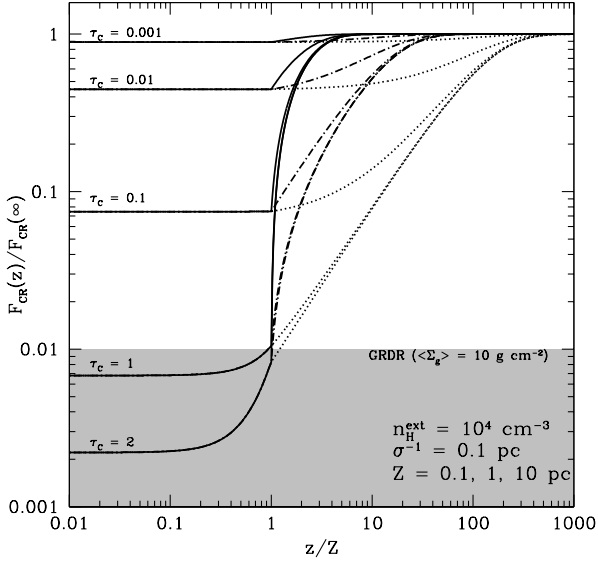}}
\figcaption[figure]{Plot of CR flux in 1D model, as a function of depth $z$, where $|z/Z| \le 1$ is the location of the overdense cloud.  The surrounding ISM is assumed to be similar to Arp 220.  Even clouds with $\tau_c \approx 0.01$ ($\sim 0.8\ \gcm2 \approx 5 \times 10^{23}\ \cm^{-2}$) have much reduced CR fluxes.  The CR flux reduction needed for gamma ray ionization to dominate is shaded in light grey for $\mean{\Sigma_g} = 10\ \gcm2$.  Cloud sizes of 0.1 (dotted), 1 (dash-dotted), and 10 pc (solid) are shown. \label{fig:CRFluxVsTau}}
\end{figure}

The slower CRs traverse a molecular cloud, the quicker that the CR flux is attenuated (assuming no CR accelerators within the cloud), and the more likely that gamma rays provide the ionization.  At best, however, CRs can traverse the cloud at $c$, with no deflection by magnetic fields within the cloud.  I consider the 1D version of this case in the Appendix, where a molecular cloud with an absorption optical depth to CRs $\tau_c$ is embedded in a medium with CR absorptivity $\alpha_{\rm out}$ and scattering coefficient $\sigma_{\rm out}$.  Far from the cloud, the CR ionization rate is $\zeta_{\rm CR}^{\rm ext}$.  In general, $\sigma_{\rm out} \gg \alpha_{\rm out}$: galaxies are scattering atmospheres for CRs.  I find that even with free travel within a molecular cloud, the CR flux is attenuated once the pionic absorption optical depth of the cloud equals $\sqrt{\alpha_{\rm out} / \sigma_{\rm out}}$ (much less than 1).  To take a specific example, the radio nuclei of Arp 220 have an average density of $10^4\ \cm^{-3}$, so the pionic lifetime of CRs within them is $\sim 5000\ \yr$ according to eqn.~\ref{eqn:tPi}.  If the (unknown) scattering mean free path within them is $\lambda_{\rm diff} = 0.1 pc$, then a cloud effectively becomes optically thick to pion production in this model when 
\begin{equation}
\Sigma_{\rm abs} \approx \sqrt{\lambda_{\rm diff} / (c t_{\pi})} \Sigma_{\pi} \approx 0.6\ \gcm2.  
\end{equation}
The rapid attenuation of CRs is shown in Figure~\ref{fig:CRFluxVsTau}.  The short effective attenuation length arises simply because CRs cross the cloud many times after being scattered by the surrounding galaxy.

\begin{figure}
\centerline{\includegraphics[width=8cm]{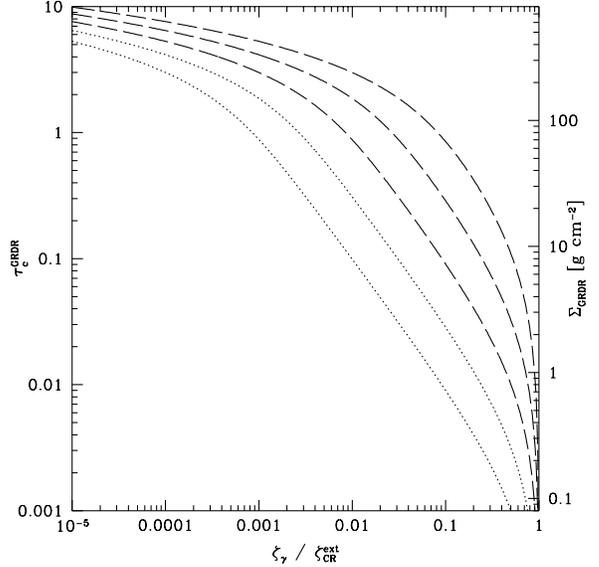}}
\figcaption[figure]{The CR absorption optical depth of a molecular cloud in which gamma-ray ionization is more important than CR ionization at its center, using the 1D model that assumes CRs free stream through the molecular cloud.  From bottom to top, the lines are for $\alpha_{\rm out} / \sigma_{\rm out} = 10^{-6}$, $10^{-5}$, $10^{-4}$, $10^{-3}$, and $10^{-2}$.\label{fig:TauGRDR}}
\end{figure}

On the other hand, it takes several optical depths before the CR ionization rate drops below the gamma-ray ionization rate.  Equation~\ref{eqn:tauGRDR} is an expression for the optical depth $\tau_{\rm GRDR}$ of a cloud that attenuates the CR flux enough so that gamma-ray ionization dominates in the center ($F(0)/F(\infty) = \zeta_{\gamma} / \zeta_{\rm CR}^{\rm ext}$).  These values are plotted in Figure~\ref{fig:TauGRDR}: I find CR absorption optical depths of $\tau_{\rm GRDR} \approx 1$ are sufficient for gamma-ray ionization to become more important than CR ionization.  Plugging in values for Arp 220's nuclei from eqn.~\ref{eqn:ZetaRatio} ($\zeta_{\rm CR}^{\rm ext} / \zeta_{\gamma} \approx 100$), I find gamma-ray ionization starts to dominate once $\tau_c \ga 0.74$.  For a typical pionic absorbing column of $80\ \gcm2$, this comes out to $\Sigma_{\rm GRDR} = 59\ \gcm2$.  A cloud with this column and density $n_c$ will have a mass $M_{\rm GRDR} \approx (4/3) \pi \Sigma_{\rm GRDR}^3/(n_c^2 m_H^2) \approx 1.5 \times 10^8\ \Msun (n_c / 10^6\ \cm^{-3})^{-2}$.  Turbulence in the molecular ISM does readily generate transient density fluctuations, but simulations show that typical column densities should still be near $\sim \mean{\Sigma_g}$ \citep{Ostriker01}, which is generally less than $10\ \gcm2$.  However, turbulence can also generate Jeans-unstable, collapsing structures such as protostars, and these can easily achieve such columns \citep{Klessen00}.  Protostars with $n_c \ga 10^{10}\ \cm^{-3}$ are very likely to be GRDRs in Arp 220's nuclei, since $M_{\rm GRDR} \approx 1.5\ \Msun$ at these densities, even if they are not shielded by a surrounding molecular cloud.

Moreover, the necessary $\tau_c$ for a GRDR grows only logarithmically with $\zeta_{\rm CR}^{\rm ext} / \zeta_{\gamma}$.  When CR scattering in the surrounding medium is slow enough ($\sqrt{\alpha_{\rm out} / \sigma_{\rm out}} \gg \zeta_{\gamma} / \zeta_{\rm CR}^{\rm ext}$), the necessary optical depth is approximately
\begin{equation}
\tau_{\rm GRDR} \approx -\ln \left[\frac{\zeta_{\gamma} / \zeta_{\rm CR}^{\rm ext}}{2 \sqrt{\alpha_{\rm out}/\sigma_{\rm out}}}\right].
\end{equation}
The larger CR lifetimes of weaker starbursts also means that CRs cross a cloud more times before being destroyed outside of the cloud: the probability of the CR being destroyed inside of the cloud rather than outside goes up.  For M82 and NGC 253, with $\zeta_{\rm CR}^{\rm ext} / \zeta_{\gamma} \approx 10000$ and a typical CR lifetime of $\sim 10^5\ \yr$, I find $\tau_c \ga 3.6$, or $\Sigma_{\rm GRDR} = 290\ \gcm2$, when $\sigma = (0.1\ \pc)^{-1}$.  Of course, gamma rays themselves start to be attenuated over such columns, but as will be seen in the next section, cascade emission from the pair $e^{\pm}$ can ensure the gamma-ray flux is preserved, to order of magnitude, out to these columns.

Of course, since these calculations are 1D, they are only illustrative.  In reality, the 1D calculations apply only for clouds that are essentially large sheets.  In three dimensions, CRs have more directions to go that do not intersect clouds.  Therefore, CRs are more likely to miss small, spherical clouds, and would cross it fewer times; the CR flux reduction need not be so drastic.

However, the above calculations are conservative in assuming that CRs can freely enter into a cloud and cross it at the speed of light.  If instead diffusion is slow inside a molecular cloud, the CR flux within it will be much smaller.  The column through which CRs can penetrate is then $\Sigma \approx \ell n m_H \approx (\sqrt{D t_{\rm life} / 3}) n m_H$.  The CRs can only diffuse into clouds with a column of less than
\begin{equation}
\Sigma \approx 0.038\ \gcm2 \left(\frac{D}{10^{28}\ \cm^2\ \sec^{-1}}\right)^{1/2} \left(\frac{n}{10^4\ \cm^{-3}}\right)^{1/2}.
\end{equation}
Thus, if CRs diffuse slowly in molecular clouds, the GRDRs may encompass large fractions of the molecular material within a starburst.  Note that hard X-rays are also not attenuated under columns $\la 2.5\ \gcm2$, so they would contribute to ionization at such low column densities.  The increase of the column with density arises because the effective speed at which CRs diffuse increases at smaller scales.  Below the mean free path, the diffusion approximation breaks down, and a treatment similar to the one above, where the CRs are assumed to free-stream, must be used.

The actual propagation of CRs in molecular clouds is poorly understood.  It is possible that CRs essentially free-stream through molecular clouds \citep{Cesarsky78,Morfill82}, which could occur if the plasma waves that normally scatter CRs are quickly damped by collisions \citep{Kulsrud71}.  However, there are reasons to expect slow diffusion in or near molecular clouds.  \citet{Skilling76} argued that because molecular clouds can absorb CRs, there is a net flux of CRs into the cloud; the net flux causes plasma instabilities in the ionized material surrounding the cloud, generating waves that scatter ionizing CRs away from the cloud.  More simply, CRs may be excluded from molecular clouds simply because the magnetic field is higher, which might lead to a smaller diffusion constant \citep[e.g.,][]{Gabici07}.  Observationally, there is radio evidence that diffusion is slow inside the Galactic Center molecular cloud Sgr B \citep{Protheroe08,Jones11}.  The diffusion constant also appears to be small in the vicinity of the molecular clouds surrounding the supernova remnants W28 and IC 443, although regions outside the molecular clouds are included in these measurements \citep[e.g.,][]{Fujita09,Gabici10,Torres10}; the small diffusion constant could be a result of plasma waves generated by the CR flux from the nearby supernova remnants in the ionized ISM \citep{Fujita10}.  \citet{Crocker11-Wild} interpret the relative gamma-ray dimness of the Galactic Center as being caused by the exclusion of CRs from molecular material, although M82 and NGC 253 seem instead to be gamma-ray bright.  

\section{The Development of the Pair Cascade: The Long Reach of Gamma Rays}
\label{sec:CascadeModel}
Gamma rays do not directly ionize gas; instead they pair produce $e^{\pm}$ that ionize the gas.  The largest uncertainty in the estimates of the GRDR ionization rate is the efficiency $f_{\rm ion}^{\gamma}$ that the energy in these pair $e^{\pm}$ is converted into ionization of the ISM.  Pair $e^{\pm}$ undergo a cascade, emitting high energy radiation, some of which may themselves pair produce $e^{\pm}$.  At low energies, ionization losses (with an energy-independent $dE/dt$) always dominate over the other losses, so low energy gamma rays are converted into ionization power relatively efficiently.  Above a few hundred MeV, bremsstrahlung losses dominate over ionization losses at all densities, since $dE/dt \propto E$ for bremsstrahlung.  Bremsstrahlung radiation from an electron of energy $E_e$ produces gamma rays with a typical energy $E_e / 2$ \citep{Schlickeiser02}.  In optically thin clouds, the gamma rays escape, so that bremsstrahlung losses waste the pair $e^{\pm}$ energy that could go into ionization.  In clouds of high enough column, though, the bremsstrahlung gamma rays can themselves pair produce $e^{\pm}$ and have another chance to ionize the ISM.  At high energy, synchrotron and Inverse Compton (IC) losses with $dE/dt \propto E^2$ are the main cooling mechanism for pair $e^{\pm}$.  The synchrotron radiation from $e^{\pm}$ of relevant energies is in the form of radio and infrared emission, which either escapes or is absorbed by dust grains.  Synchrotron emission therefore does not contribute to the ionization of GRDRs.  IC radiation is ultraviolet light, X-rays, and gamma rays.  Ultraviolet and X-rays can directly ionize atoms.  As with bremsstrahlung, gamma rays either escape or pair produce depending on the cloud column density.

Estimating $f_{\rm ion}^{\gamma}$ requires modeling of all of these cooling processes.  Furthermore, the gamma-ray spectral features of starburst galaxies must be taken into account.  The gamma rays of starbursts are thought to be mostly pionic; because of the kinematic threshold of pion production, the pionic spectrum should drop away at energies below 70 MeV in $dN/dE$ \citep[e.g.,][]{Stecker70}, or a few hundred MeV in $E^2 dN/dE$ (for example, $\sim 700-800\ \MeV$ for the Galactic pionic spectrum, as in \citealt{Kamae06}).  At high energies, the spectrum will be a hard power law.  As a result, most of the pair $e^{\pm}$ will be at high energy, where they will not directly ionize the gas but instead cool by radiation.  Estimating the ionizing effects of the resulting gamma-ray cascade is the subject of this section.

I make many simplifying assumptions.  Once again, I use a 1D two stream model for the gamma rays.  In this case, I assume that $e^{\pm}$ neither diffuse nor are advected from the regions they are created (an on-the-spot approximation), which would be the case if $D \to 0$ in the in the dense environments of molecular clouds.\footnote{By contrast, calculations of ionization in protostars and protoplanetary disks assume that CRs stream at $c$ \citep[e.g.,][]{Umebayashi81}.  While this may be appropriate in these environments, over the much larger scales of a molecular cloud, magnetic deflection can be important.}  I can phrase the equation of radiative transfer in terms of the column $N_H = n_H z$ instead of the physical depth $z$.  Each generation $i$ of gamma rays produces pair $e^{\pm}$, which in turn generate the gamma rays of the next generation $i + 1$.  The equation of radiative transfer for these assumptions is:
\begin{equation}
\frac{dI_{\pm}^i}{dN_H} = \pm \left(-\sigma_{\gamma Z} I_{\pm}^i + \frac{Q_{\rm brems}^{i - 1} + Q_{\rm IC}^{i - 1}}{2 n_H}\right),
\end{equation}
where for the primary gamma rays ($i = 1$), $Q_{\rm brems}^{i - 1} = Q_{\rm IC}^{i - 1} = 0$, and $\sigma_{\gamma Z} = \alpha_{\gamma Z} / n_H$ is the total cross section for $\gamma Z$ pair production.  The intensity $I_{\pm}^i$ is a number intensity per unit energy here rather than energy intensity.  The 1/2 factor for the $Q$ terms takes into account that half of the emitted radiation would be emitted towards higher columns (in $I_+$), and half towards lower columns (in $I_-$), since the direction of $e^{\pm}$ are likely to be scrambled by magnetic fields.  

Pair production and the radiation from the pairs is calculated using $\delta$-function approximations.  For $\gamma Z$ pair production, the pair $e^{\pm}$ from a photon of energy $E_{\gamma}$ are assumed to all be of energy $E_e = E_{\gamma} / 2$:
\begin{equation}
Q_e^i = \displaystyle 4 \alpha_{\gamma Z} (2E_e) \int I^i (2 E_e) d\Omega  = 4 (I_+^i (2 E_e) + I_-^i (2 E_e)) \alpha_{\gamma Z} (2E_e).
\end{equation}
I then estimate the pair $e^{\pm}$ spectrum as 
\begin{equation}
N_e^i = Q_e^i \times t_{\rm life} 
\end{equation}
where $t_{\rm life} = [t_{\rm ion}^{-1} + t_{\rm brems}^{-1} + t_{\rm synch}^{-1} + t_{\rm IC}^{-1}]^{-1}$ includes losses from ionization, bremsstrahlung, synchrotron, and Inverse Compton.  Defining the loss times for each process as $E / (-dE/dt)$, I use the Bethe-Bloch energy losses for ionization from \citet{Strong98}, the bremsstrahlung losses in a neutral hydrogen gas from \citet{Strong98}, and the standard Thomson energy loss formulas for synchrotron and IC emission from \citet{Rybicki79}.

Bremsstrahlung emission is assumed to come out at energies $E_{\gamma} = E_e / 2$:
\begin{equation}
Q_{\rm brems}^{i + 1} = \frac{4 N_e^i (2E_{\gamma})}{t_{\rm brems}}.
\end{equation}
Finally IC radiation is assumed to come out at an energy $E_{\gamma} = 4/3 (E_e / m_e c^2)^2 \epsilon_0$ \citep{Rybicki79}, where $\epsilon_0 = 2.7 kT$ is the energy of photons in the background radiation field.  Since IC spreads the energy of one dex in $e^{\pm}$ energy into two dex of upscattered photon energy, I have $E_{\gamma}^2 Q_{\rm IC}^i = E_e^2 N_e / (2 t_{\rm IC})$:
\begin{equation}
Q_{\rm IC}^{i + 1} = \frac{E_e^2 N_e^i (E_e)}{2 E_{\gamma}^2 t_{\rm IC}}.
\end{equation}

The volumetric energy input in ionization losses is calculated as
\begin{equation}
\Gamma_{\rm ion} = \int \frac{E_e N_e^i}{t_{\rm ion}} dE_e.
\end{equation}
In addition, I assume that Inverse Compton photons with energies between 13.6 eV and 1 MeV are immediately absorbed, and contribute all of their energy to ionization.  EUV and soft X-ray ($\la 10\ \keV$) photons can directly ionize neutral atoms through photoabsorption, though in practice, they can also heat the ISM by exciting atoms without ionizing it.  Hard X-rays lose energy through Compton scattering in large column depths ($\ga 10^{24}\ \cm^{-2}$).  The energy lost per scattering is hundreds of eV to many keV, so these scatterings eject electrons from atoms, which then can act as secondary electrons to ionize other atoms.  I calculate the volumetric low energy IC energy input as
\begin{equation}
\Gamma_{\rm LE-IC} = \int_{\rm 13.6\ eV}^{\rm 1\ MeV} E_{\gamma} Q_{\rm IC} dE_{\gamma}.
\end{equation} 
Then the ionization rate is just
\begin{equation}
\xi = \frac{\Gamma_{\rm LE-IC} + \Gamma_{\rm ion}}{n_H E_{\rm ion}}.
\end{equation}

I applied this method to an input gamma-ray spectrum from the fiducial models of \citet{Lacki10-XRay} of M82, which is detected in gamma rays, and Arp 220's east nucleus.  The western nucleus of Arp 220 would likely provide roughly similar results \citep[c.f.][]{Torres04}, but it is unclear how much it is powered by an active galactic nucleus \citep[e.g.,][]{Downes07}.  The input gamma-ray spectrum is in terms of unabsorbed volumetric power emitted $\epsilon$; I convert it to an intensity by multiplying by a midplane-to-edge scale height of $h$, set to 100 pc for M82 and 50 pc for Arp 220's east nucleus: $I_+^1 (N_H = 0) = \epsilon h$.  I assume the modeled column is completely opaque, so that $I_-^1 (N_H^{\rm max}) = 0$.  The spectrum extends down to 1 MeV, and I cut it off at 10 TeV, since higher energy gamma rays are likely absorbed by $\gamma\gamma$ pair production processes \citep{Torres04,Inoue11,Lacki10-XRay}.  I adopt a density of $10^6\ \cm^{-3}$ and a magnetic field strength of 10 milliGauss, similar to those found by Zeeman splitting measurements of dense OH masers \citep{Robishaw08}.  The radiation field has a blackbody spectrum with temperature of 40 K for M82 and 50 K for Arp 220, but is scaled to a characteristic radiation energy density $U_{\rm rad}$.  M82 is optically thin in the far-infrared, so I calculate $U_{\rm rad} = L_{\rm IR} / (2 \pi R^2 c)$, where $L_{\rm IR} = 5.9 \times 10^{10}\ \Lsun$ is the total IR luminosity from \citet{Sanders03} and $R = 250\ \pc$ is the radius.  Arp 220's nuclei are optically thick in the far-infrared \citep[e.g.,][]{Downes07,Sakamoto08,Papadopoulos10-A220}, so I just use the blackbody energy density.

\begin{figure*}
\centerline{\includegraphics[width=9cm]{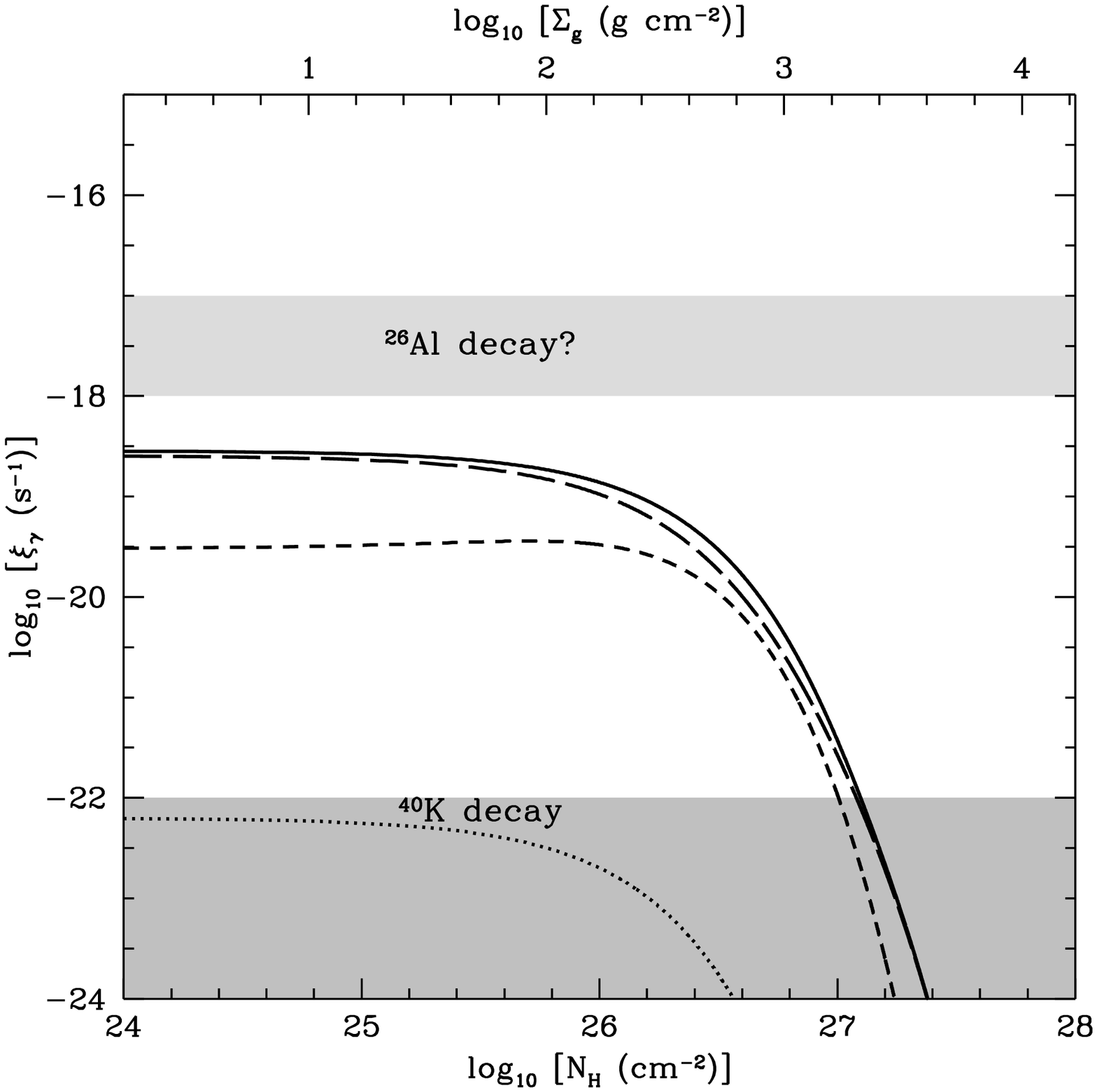}\includegraphics[width=9cm]{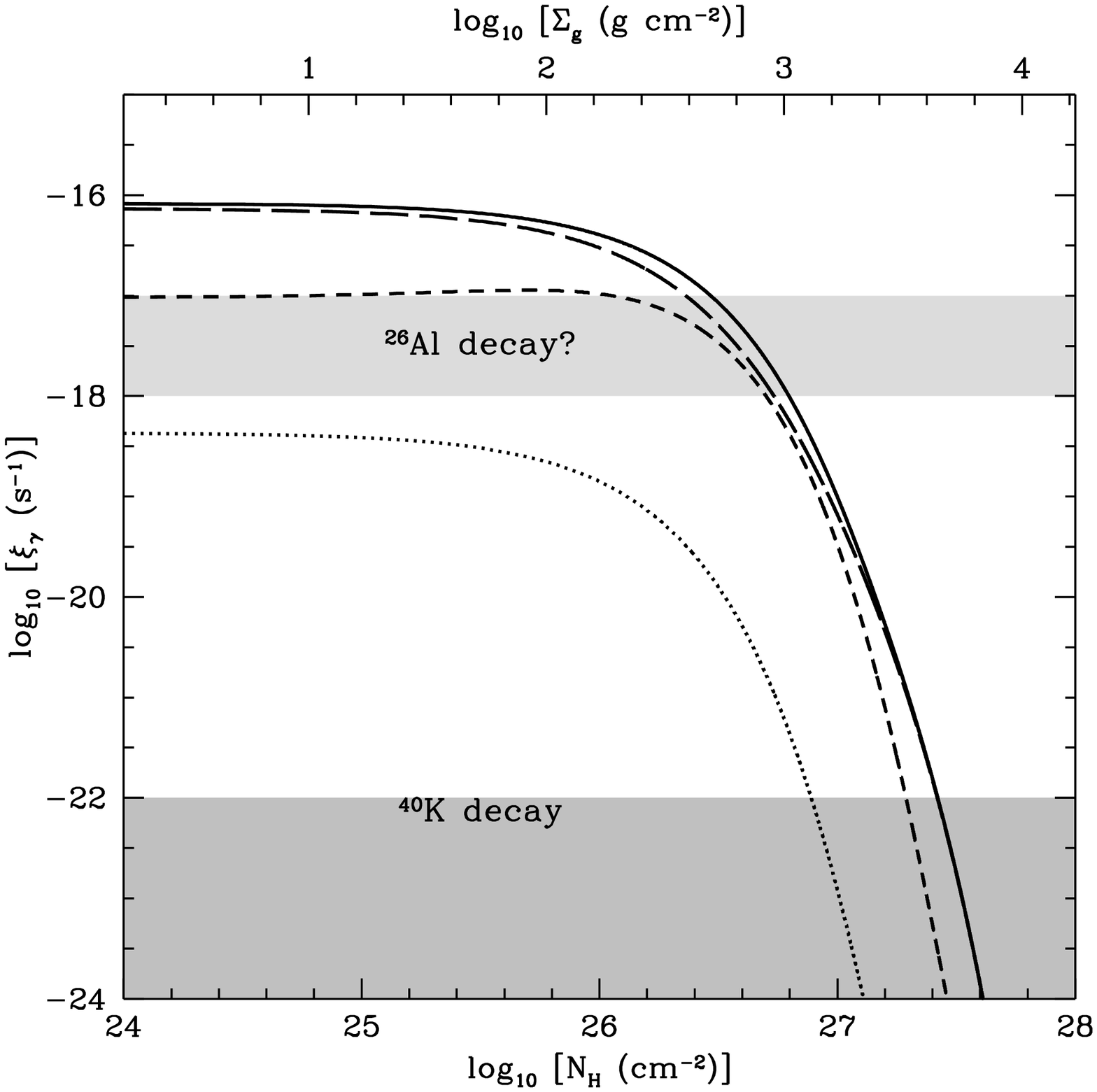}}
\figcaption[figure]{Plot of ionization rate using a prediction for the gamma-ray spectrum of M82 (left) and Arp 220's east nucleus (right), using a 1D model and assuming $n_H = 10^6\ \cm^{-3}$.  Cascade $e^{\pm}$ are assumed to remain in place ($D = 0$).  I assume that the power in all Inverse Compton radiation between 13.6 eV and 1 MeV instantly ionizes the surrounding gas (dotted line).  Total ionization rate (solid), ionization rate from primary gamma-rays (long-dashed line), and ionization rate from cascade gamma-rays (short-dashed line) are plotted.  \label{fig:Arp220EIonization}}
\end{figure*}

The resulting ionization rates for a large column of material are displayed in Figure~\ref{fig:Arp220EIonization}.  I find that, to order of magnitude, the gamma-ray ionization rate at low column depths is similar to that derived in section~\ref{sec:Estimates} using $f_{\rm ion}^{\gamma} \approx 0.1$.  In M82, $\zeta_{\gamma} \approx 3 \times 10^{-19}\ \sec$, about three times higher than predicted; in Arp 220 East, $\zeta_{\gamma}$ is about $8 \times 10^{-17}\ \sec^{-1}$, about twice the estimated value.  In reality, these values will vary for GRDRs with different densities, magnetic field strengths, and radiation fields.  

In both M82 and Arp 220 East, the ionization rate is maintained to enormous column densities.  In the M82 GRDR, the ionization rate remains above $10^{-19}\ \sec^{-1}$ to $240\ \gcm2$ ($N_H = 1.5 \times 10^{26}\ \cm^{-2}$), above $10^{-20}\ \sec^{-1}$ through $790\ \gcm2$ ($N_H = 4.7 \times 10^{26}\ \cm^{-2}$), and above the $^{40}$K radioactivity-induced ionization rate of $10^{-22}\ \sec^{-1}$ through $2100\ \gcm2$ ($N_H = 1.3 \times 10^{27}\ \cm^{-2}$).  I find that in the Arp 220 East GRDR, gamma rays are capable of sustaining ionization rates $\ge 5 \times 10^{-17}\ \sec^{-1}$, the canonical value in Milky Way molecular clouds, to columns of $7.0 \times 10^{25}\ \cm^{-2}$ ($117\ \gcm2$). Even at columns of $2.9 \times 10^{26}\ \cm^{-2}$ ($490\ \gcm2$), the ionization rate is still $\sim 10^{-17}\ \sec^{-1}$, and it drops to $10^{-18}\ \sec^{-1}$ only after $6.3 \times 10^{26}\ \cm^{-2}$ ($1050\ \gcm2$).  Thus, even if short-lived radioisotopes like $^{26}$Al are well-mixed with the molecular gas, sustaining an ionization rate of $10^{-18} - 10^{-17}\ \sec^{-1}$ on their own \citep{Lacki12-Al26}, gamma-rays will be the dominant ionization source to columns of several hundred $\gcm2$ in Arp 220 East.  It takes a column of $4400\ \gcm2$ ($N_H = 2.6 \times 10^{27}\ \cm^{-2}$) to reduce $\zeta_{\gamma}$ to less than the $^{40}$K radioactivity induced rate of $10^{-22}\ \sec^{-1}$ in Arp 220 East.  The reach of GRDRs is extended somewhat by high energy cascade $e^{\pm}$, which can regenerate gamma rays through bremsstrahlung and IC emission (short-dashed line).

I conclude that $f_{\rm ion}^{\gamma} \approx 0.1$ is a useful approximation, that gamma rays can easily provide significant ionization through columns of $100\ \gcm2$, and even with attenuation, they may be the dominant source of ionization through columns of several thousand $\gcm2$.

\section{Observational Prospects}
\label{sec:Observations}
High energy gamma rays ionize material through the pair $e^{\pm}$ they produce: the chemistry of gamma-ray ionization will therefore be the same as for CR $e^{\pm}$ ionization, and CR ionization in general.  The chemical signatures of cosmic ray ionization have been considered recently in several recent papers in the context of CRDRs \citep{Meijerink06,Meijerink11,Bayet11}.  The advent of ALMA, SOFIA, and \emph{Herschel} allows relevant lines, such as high-$J$ transitions of CO, to be measured.  GRDRs probe a different regime of ionization and temperature not considered by these papers, and might have distinct chemistry.  Unlike CRDRs where the dust temperature is less than the gas temperature, I have argued that the gas temperature in GRDRs is less than or equal to the dust temperature.  The gamma-ray ionization rate of GRDRs may make them more similar to typical PDRs, but unlike PDRs, the columns in GRDRs are so high that there is essentially no UV light whatsoever.    

One feature distinguishing GRDRs from CRDRs is that GRDRs should be cold (section~\ref{sec:GRDRTemperature}).  Whereas CRDRs will have gas temperatures of up to $50 - 150\ \Kelv$, GRDRs will have minimum gas temperatures $\la 10\ \Kelv$ even in Arp 220 without dust heating, and maximum gas temperatures equal to the dust temperature.  The GRDR gas temperatures could be much smaller in less extreme starbursts, $\la 1\ \Kelv$, depending on dust-gas interactions and turbulent heating.  The presence of cold, dense gas would therefore support the idea that there are regions CRs cannot penetrate into.  

However, a severe difficulty with actually constraining GRDR properties with observations arises because of their high column densities.  The optical depth of a column of gas can be parameterized as
\begin{equation}
\tau_{\rm dust} = N_H (\tau_0 / N_0) \left(\frac{\lambda}{\lambda_0}\right)^{-\beta}.
\end{equation}
Planck observations of local Galactic molecular clouds find that $\tau_0 / N_0 \approx 2.3 \times 10^{-25}\ \cm^2$ at $\lambda = 250\ \mu m$ in molecular gas, assuming $\beta = 1.8$ \citep{Abergel11}.  This means that the optical depth is unity out to 
\begin{equation}
\label{eqn:GRDRDustObscuration}
\lambda_1 \approx 1.1\ \mm \left(\frac{\Sigma_g}{100\ \gcm2}\right)^{0.55}.
\end{equation}
Indeed, the west nucleus of Arp 220 is, if anything, even worse, being optically thick out to millimeter wavelengths with $\mean{\Sigma_g} \approx 10\ \gcm2$ ($\mean{N_H} \approx 10^{25}\ \cm^{-2}$) \citep[e.g.,][]{Downes07,Sakamoto08,Papadopoulos10-A220}.  If CRs diffuse very slowly in dense starburst molecular gas, then gamma-ray ionization may dominate for cloud columns as small as $0.01\ \gcm2$ (section \ref {sec:ScatteringEffect}), so that GRDRs can be optically thin for wavelengths longer than $\sim 6\ \micron$, though X-ray ionization can still be important through these columns.  However, if CRs free-stream through GRDR clouds, then gamma-ray ionization should only be important at columns $\sim 100\ \gcm2$.  In addition to the dust continuum opacity, molecular lines such as CO will have additional opacity.  

If GRDRs are really so optically thick, then, their defining feature will be their shadows in molecular lines against the brighter, hotter CR-ionized material behind them.  Because GRDR gas is colder than the CRDR gas, and if GRDRs are preferentially in dense cores with low levels of temperature, the molecular lines should have different kinematics than the CRDR gas, with lower turbulent and thermal line widths, making them easier to distinguish.   If CRs instead pervade the entire starburst, they will heat all of the gas to the same temperature, and no molecular line shadows should be visible.  These molecular line shadows may be detectable in the nearest starburst galaxies with ALMA, which will be able to make detailed and accurate images with an angular resolution of $\sim 0\farcs1$ \citep{Brown04}.  At the distance of M82 ($\sim 3.6\ \Mpc$), this translates to 1.7 pc of spatial resolution.  A cloud of this radius and density of $10^6\ \cm^{-3}$ has a column density of $\Sigma_g = 9\ \gcm2$ ($N_H = 5 \times 10^{24}\ \cm^{-2}$; mass of $5 \times 10^5\ \Msun$), large enough to be an interesting test of whether CR diffusion is fast or slow in starburst molecular clouds.  The dust in GRDRs will have the same temperature as that outside of them, so there are no similar shadows in dust continuum emission.

A final possibility for directly testing gamma-ray ionization is to look for signs of gamma-ray pair production in the radio, by looking for synchrotron emission from the $\gamma Z$ pair $e^{\pm}$.  There have been studies looking in the Milky Way for CR penetration into cores by searching for synchrotron from pionic secondary $e^{\pm}$ \citep[e.g.,][]{Jones08,Jones11}.  The competing signal is the diffuse radio synchrotron emission from the starburst.  In starbursts, the diffuse $e^{\pm}$ population is likely dominated by secondary $e^{\pm}$ produced by the same pionic loss process that produces most of the gamma rays \citep[e.g.,][]{Rengarajan05,Lacki10-FRC1}.  It is therefore useful to relate both the diffuse synchrotron and synchrotron from GRDRs to the gamma-ray emission of the starburst.  For GRDRs with optical depth $\tau_{\gamma Z} \la 1$, the volumetric synchrotron power is related to the incident gamma-ray flux $\nu F_{\nu} (\GeV)$ as $\nu \epsilon_{\nu}^{\rm synch} ({\rm GRDR}) \approx (1/2) \nu F_{\nu} (\GeV) \tau_{\gamma Z} f_{\rm synch}^{\gamma} / (2 \ell)$, where $f_{\rm synch}^{\gamma}$ is the efficiency that gamma-rays are converted into synchrotron radiation.  The factor of $1/2$ arises because the synchrotron frequency depends on the electron energy (and primary gamma-ray energy) squared.  Setting $\nu F_{\nu} (\GeV) \approx 4\pi \nu I_{\nu} (\GeV)$ and $\nu I_{\nu}^{\rm synch} ({\rm GRDR}) = \nu \epsilon_{\nu}^{\rm synch}  ({\rm GRDR}) \ell / (4\pi)$, I find
\begin{equation}
\frac{\nu I_{\nu}^{\rm synch} ({\rm GRDR})}{\nu I_{\nu} (\GeV)} \approx \tau_{\gamma Z} f_{\rm synch}^{\gamma} / 2.
\end{equation}
For the diffuse synchrotron and gamma-ray emission from starbursts, the ratio of GHz synchrotron to GeV luminosity is $\sim 20 - 40$ \citep{Lacki11-Obs}.  If $f_{\rm synch}^{\gamma} \approx 1$, the ratio can be up to $\sim 1/2$ for GRDRs.  Assuming $\nu I_{\nu} (\GeV)$ within the GRDR is equal to its mean value throughout the starburst, the potentially larger synchrotron radiation efficiency means that the radio intensity towards a GRDR can be up to $\sim 10 - 20$ times larger than on sightlines through the diffuse starburst.  However, this depends on how pair $e^{\pm}$ cool in the GRDR; if $f_{\rm synch}^{\gamma} \la 0.1$, then the radio intensity on the GRDR sightline will not be enhanced much.  Furthermore, it may be difficult in practice to distinguish pair $e^{\pm}$ made by gamma rays from pionic $e^{\pm}$ made by CRs, although synchrotron emission from a cloud would still be a sign that \emph{some} kind of high-energy particle is penetrating into it.

It is therefore difficult to directly test for the presence of GRDR ionization, because GRDRs have low ionization rates in the first place, they have low heating rates, and are probably opaque to infrared lines that could inform us about the chemistry in GRDRs.  We may only be able to discover cold regions where CRs do not penetrate and note that in gamma-ray detected starbursts like M82, gamma-ray ionization must exist.  Fortunately, gamma rays are not deflected by magnetic fields, unlike CRs, so there is little question that gamma rays can penetrate into molecular clouds, though there remain uncertainties in $f_{\rm ion}^{\gamma}$, because of the intermediate step of the cooling of pair $e^{\pm}$.

An indirect method of studying GRDRs is through their effect on star-formation, particularly the Initial Mass Function (IMF).  If starburst star-forming regions are CRDRs, the gas should be heated to temperatures of $\sim 100\ \Kelv$, raising the Jeans mass and forcing the IMF to be top-heavy \citep{Papadopoulos11-SF}.  However, GRDRs are much colder, resulting in much smaller Jeans masses, perhaps even smaller than in the Milky Way since the gas temperatures in GRDRs can be less than 1 K at low densities (section~\ref{sec:GRDRTemperature}).  As I showed in section~\ref{sec:GRDRTemperature}, the GRDR Jeans mass is less than $2\ \Msun$ for all densities, and is roughly $1\ \Msun$ for densities of $10^4\ \cm^{-3}$ and $10^6\ \cm^{-3}$.  Observations of the IMF of stellar populations formed in starbursts should constrain whether star-formation occurs primarily in warm CRDRs, cold GRDRs, or some mix of the two.  There have been conflicting results about what the IMF in starbursts is; some studies have suggested that starburst environments have top-heavy IMFs \citep[e.g.,][]{Rieke80,Figer99,Smith01,McCrady03}, but others are consistent with a normal IMF in starbursts (e.g., \citealt{McCrady03,Kim06,Tacconi08}; see also the review by \citealt{Bastian10}).  On the other hand, \citet{vanDokkum10} recently argued that elliptical galaxies have a bottom-heavy IMF (see also \citealt{vanDokkum11}).  If confirmed, the evidence for a bottom-heavy IMF would support the idea that collapsing gas in starbursts passes through a GRDR phase cold enough to fragment into small mass protostars.  

\section{Conclusions}
\label{sec:Conclusions}
Radiation ionizes media with $\tau \approx 1$ most effectively.  At low optical depth, the radiation simply escapes; at high optical depth, the radiation ionizes the outer layers of the medium, but the interior is shielded.  In order to guarantee ionization of the entire ISM, different kinds of ionizing radiation are needed with different penetration depths.  In starburst galaxies, even CRs may be destroyed by dense molecular gas before they can ionize and heat molecular gas at the highest columns.  I have argued that gamma rays, produced by the interaction of CRs with that gas, can provide a guaranteed ionization rate through columns of $100\ \gcm2$ or more.  

I have calculated the basic properties of the resulting GRDRs:
\begin{itemize}
\item The GRDR ionization rate is small, expected to be of order $10^{-19}\ \sec^{-1}$ in M82 and $10^{-16}\ \sec^{-1}$ in Arp 220's radio nuclei (Figure~\ref{fig:GRDRZeta}).  The latter is comparable to the Milky Way molecular cloud ionization rate.  However, gamma-ray ionization still is more important than ionization from long-lived radioactive nuclei such as $^{40}$K in the ISM for most starbursts.

\item The ionization fraction of GRDRs is small, $\sim 10^{-11} - 10^{-8}$ for $n_H = 10^6\ \cm^{-3}$ (Figure~\ref{fig:GRDRXe}).  But even these small ionization fractions are sufficient to slow down ambipolar diffusion to timescales comparable to or slower than the free-fall time of gas in starbursts as dense as M82, if the \citet{McKee89} formulas hold in such environments.  This helps dense gas in starbursts retain magnetic fields, suggesting star formation proceeds similarly as in dense cores in the Milky Way.  On the other hand, the ambipolar diffusion time is less than 100 times the gas free-fall time even in Arp 220, which may solve the problem of ambipolar diffusion halting star-formation in CRDRs.

\item GRDRs are very cold, with minimum gas temperatures of $\la 10\ \Kelv$ even for Arp 220-like starbursts, and $\la 1\ \Kelv$ in starbursts like M82 (Figure~\ref{fig:GRDRTemp}).  Most likely, the heating of GRDRs comes from dust-gas interactions or possibly turbulence rather than gamma-ray heating, raising their gas temperature to the ambient dust temperature.

\item The reason the GRDR ionization rate is so much smaller than the CRDR ionization rate (Figure~\ref{fig:GRDRZeta}) is that CRs are scattered many times in starbursts, while gamma rays pass freely through them.  The CR flux through clouds of low optical depth is therefore increased by a large factor.  However, this poses a disadvantage for CR ionization at large columns: the column that CRs must penetrate is much higher than that for gamma rays.  As a result, CRs may be depleted and gamma-ray ionization may take over in clouds of column $\la 100\ \gcm2$ in Arp 220 (see Figures~\ref{fig:CRFluxVsTau} and~\ref{fig:TauGRDR} for the 1D case).  If CR diffusion is slow, even clouds with column as small as $0.01\ \gcm2$ may be GRDRs, depending on the presence of X-rays.

\item Calculations of the actual ionization rate $\zeta_{\gamma}$ in GRDRs are complicated by the cascade that develops from $\gamma Z$ pair production.  I find that assuming $f_{\rm ion}^{\gamma} = 10\%$ as an ionization efficiency works as a rough guide to the ionization rate in densities of $10^6\ \cm^{-3}$ (section~\ref{sec:CascadeModel}).  The gamma-ray ionization rate does not appreciably drop until columns of $100\ \gcm2$.  Gamma-ray ionization can dominate over $^{40}$K radioactivity ($\zeta_{\gamma} \ga 10^{-22}\ \gcm2$) through columns of several thousand $\gcm2$.  

\item Observations of GRDRs are likely to be difficult, because they occur in regions of columns high enough to obscure even the infrared and submillimeter (eqn.~\ref{eqn:GRDRDustObscuration}).  In principle, the chemical signatures should be similar to low levels of cosmic ray ionization.  Because they are relatively cold, GRDRs would appear as shadows in molecular lines against the hotter CRDR regions behind them, and would have different line kinematics than the background CRDRs.  These molecular line shadows may be visible in nearby starbursts like M82 with ALMA.  Synchrotron emission from the pair $e^{\pm}$ in GRDRs may also be visible in the radio, though it is unclear if such observations could distinguish gamma-ray produced $e^{\pm}$ from pionic $e^{\pm}$ made by CR protons.

\item The cold gas of GRDRs means that their Jeans masses can be very small (equation~\ref{eqn:MJeansNumer}).  If star-forming gas passes through a GRDR phase, it can fragment on small scales, and may have a normal or even bottom-heavy IMF.  This is in contrast to CRDRs, which are much hotter and must have a top-heavy IMF.
\end{itemize}

While my focus in this paper has been on gamma rays from starbursts, GRDRs can also be present around AGNs.  Indeed, while gamma ray production in starbursts is limited by the relatively small fraction of bolometric luminosity that goes into CRs, these limits do not apply to AGNs.  Furthermore, the inverse square dependence of the flux on the distance \citep[c.f.,][]{Papadopoulos10-CRDRs} actually helps for gas near AGNs.  Therefore, the gamma-ray ionization rate near AGNs may be much higher, and GRDRs may set the conditions for star formation that appears to happen near gamma-ray bright AGNs \citep{Paumard06,Davies07}.

GRDRs are the next rung in the hierarchy of ionization regions after CRDRs (Figure~\ref{fig:Explanation}).  It is natural to ask whether anything is even more penetrating than gamma rays.  One known source of ionization at arbitrarily high columns is radioactivity from unstable isotopes in the ISM, such as $^{40}$K \citep{Cameron62,Umebayashi81}.  The $^{26}$Al produced by the many young stars in starbursts may increase the radioactive ionization rate of the ISM further \citep{Umebayashi09}.  If $^{26}$Al is rapidly mixed with the molecular gas of starbursts, it should sustain ionization rates of $10^{-18} - 10^{-17}\ \sec^{-1}$ \citep{Lacki12-Al26}.  Then $^{26}$Al would dominate the ionization of dense molecular gas in weaker starbursts (including M82) and gamma rays would dominate the ionization rate in stronger starbursts (like Arp 220's nuclei), as shown in Figure~\ref{fig:GRDRZeta}.  Even if $^{26}$Al does not contribute to the ionization rate, the gamma rays provide a relatively certain source of ionization in extremely dense gas.  As noted in section~\ref{sec:CascadeModel}, gamma rays can provide greater ionizing rates than $^{40}$K radioactivity even to column depths of several thousand $\gcm2$ in starbursts.   

This would be very relevant for protostars and protoplanetary disks, which can achieve column densities of $1000\ \gcm2$.  It is thought that protoplanetary disks in the Milky Way have ``dead zones'' where there is essentially no ionization at column densities greater than $100\ \gcm2$.  In the dead zone, the gas is no longer affected by magnetic fields and accretion may be halted \citep[e.g.,][]{Gammie96}.  According to \citet{Gammie96}, dead zones appear in regions with $x_e \la 10^{-13}$; for regions of characteristic density of $n_H = 10^{13}\ \cm^{-3}$, this corresponds to an ionization rate of $\zeta \approx 5 \times 10^{-20}\ \sec^{-1}$, smaller than the gamma-ray ionization rate in Arp 220 by several orders of magnitude.  Hence, in starbursts the dead zones may be reduced or even non-existent because of gamma-ray ionization, which may affect the process of planet formation (c.f. \citealt{Fatuzzo06}, which discusses the effects of high CR ionization on protoplanetary disks).  Another possible region where gamma rays would be attenuated are nuclear torii surrounding AGNs, which can have clumps with column densities approaching $N_H = 10^{27}\ \cm^{-2}$ ($\Sigma_g = 1700\ \gcm2$) \citep[][and references therein]{Hopkins12}.

A more speculative radiation source is dark matter annihilation or decay.  Dark matter is distributed throughout galaxies and is not stopped by matter of any reasonable density, although it would be concentrated more towards the galactic center, where any active nucleus would be.  High concentrations of dark matter may produce high energy radiation which will cascade down and contribute to ionization.  The most extreme regions of starbursts or active nuclei would then be Dark Matter Dominated Regions (DMDRs), and indeed the entire early Universe before reionization may have been a giant DMDR, when there was no other source of ionizing radiation \citep[e.g.,][]{Padmanabhan05,Mapelli06}.

\acknowledgments
I would like to acknowledge inspiring conversations with and comments from Padelis Papadopoulos.  I extend a general thanks to the participants of the ``Cosmic Ray Interactions: Bridging High and Low Energy Astrophysics'' conference, which first motivated this line of thought for me.  I am grateful to Todd Thompson who discussed planet formation in high ionization environments with me, and provided useful comments on the paper.  I would also like to thank Diego Torres for discussions on CR diffusion.  I was supported by a Jansky Fellowship for this work from the NRAO.  NRAO is operated by Associated Universities, Inc., under cooperative agreement with the National Science Foundation.

\appendix
I consider 1D radiative transfer in a proton calorimetric starburst.  For the 1D problem, there are just two sightlines, one extending to $+z$ with an intensity $I_+$ and the other extending to $-z$ with an intensity $I_-$.  The equation of radiative transfer is
\begin{equation}
\label{eqn:1DRadTransfer}
\frac{dI_{\pm}}{dz} = \pm \left(\frac{Q(z)}{2} - \alpha(z) I_{\pm} - \sigma(z) (I_{\pm} - J)\right)
\end{equation}
where $J = (I_+ + I_-)/2$ is the angle-averaged intensity at that $z$, $\alpha(z)$ is the inverse of the mean free path for CR destruction (absorption coefficient), and $\sigma (z)$ is the inverse of the mean free path for CR scattering, and $Q (z)$ is the rate that CRs are injected into the ISM \citep{Rybicki79}.  Suppose a molecular cloud sits at $-Z < z < Z$, through which CRs free stream at $c$.  Outside of the cloud, CRs can be either scattered or absorbed; I assume this surrounding, large-scale medium is homogeneous.  The surrounding starburst is also assumed here to be infinitely large, which should yield the correct behavior as long as the starburst is large compared to the typical diffusion length of protons.  The cloud, being denser than the surrounding medium has a different absorption coefficient than the surrounding medium.  Furthermore, I assume that no CRs are emitted inside the cloud.  Therefore: 
\begin{eqnarray}
\alpha(z) & \approx & \left\{ \begin{array}{ll} \alpha_c & (z < Z)\\ \alpha_{\rm out} & (z > Z) \end{array} \right.\\
\sigma(z) & \approx & \left\{ \begin{array}{ll} 0 & (z < Z)\\ \sigma_{\rm out} & (z > Z) \end{array} \right.\\
Q(z) & \approx & \left\{ \begin{array}{ll} 0 & (z < Z)\\ Q_{\rm out} & (z > Z) \end{array} \right.
\end{eqnarray}
I define the CR absorption optical depth from the edge of the cloud to its center ($z = Z$ to $z = 0$) as $\tau_c = \alpha_c Z$.

The equation can be solved for $z \ge Z$ by differentiating both sides by $z$:
\begin{equation}
\frac{d^2 I_+}{dz^2} = -(\alpha_{\rm out} + \sigma_{\rm out}/2) \frac{dI_+}{dz} + \sigma_{\rm out}/2 \frac{dI_-}{dz}
\end{equation}
and then using \ref{eqn:1DRadTransfer} to find, in terms of $I_+$, substitutions for $dI_-/dz$ and then $I_- = 2/\sigma_{\rm out} [dI_+/dz - Q_{\rm out}/2 + (\alpha_{\rm out} + \sigma_{\rm out}/2) I_+]$.  The final differential equation is then
\begin{equation}
I_+ = \frac{1}{\alpha_{\rm out} (\alpha_{\rm out} + \sigma_{\rm out})} \frac{d^2 I_+}{dz^2} + \frac{Q_{\rm out}}{2 \alpha_{\rm out}}.
\end{equation}
After requiring $I_{\pm}$ to be finite as $z \to \infty$, the solution to the differential equation for $z \ge Z$ is:
\begin{eqnarray}
I_+ & = & \displaystyle C e^{-\alpha_{\rm out} \mu z} + \frac{Q_{\rm out}}{2\alpha_{\rm out}}\\
I_- & = & \displaystyle 2 \frac{C}{\sigma_{\rm out}} e^{-\alpha_{\rm out} \mu z} (\alpha_{\rm out} + \sigma_{\rm out}/2 - \alpha_{\rm out}\mu) + \frac{Q_{\rm out}}{2\alpha_{\rm out}},
\end{eqnarray}
where I define $\mu \equiv \sqrt{1 + \sigma_{\rm out}/\alpha_{\rm out}}$.  In these equations, $C$ is a constant set by the boundary conditions at the edge of the molecular cloud ($z = Z$).  From symmetry, the intensity of CRs entering into the cloud at $z = +Z$ must equal the intensity of CRs entering into the cloud at $z = -Z$, so that $I_- (Z) = I_+ (-Z)$.  In addition, since there is no CR scattering within the cloud, the CR intensity emerging out of the cloud is just the intensity emerging into the cloud attenuated by absorption: $I_+(Z) = I_-(Z) e^{-2\alpha_c Z}$ and likewise $I_-(-Z) = I_+(-Z) e^{-2 \alpha_c Z}$.  To find $C$, I therefore impose the additional condition that $I_+(Z)= I_-(Z) e^{-2\tau_c}$.  This gives me for $z \ge Z$
\begin{eqnarray}
I_+ & = & \displaystyle \frac{Q_{\rm out}}{2\alpha_{\rm out}} \left[e^{\alpha_{\rm out}\mu(Z - z)} \frac{e^{-2\tau_c} - 1}{1 - (2/\sigma_{\rm out})(\alpha_{\rm out} + \sigma_{\rm out}/2 - \alpha_{\rm out}\mu)e^{-2\tau_c}} + 1\right]\\
I_- & = & \displaystyle \frac{Q_{\rm out}}{2\alpha_{\rm out}} \left[\frac{2}{\sigma_{\rm out}} e^{\alpha_{\rm out}\mu(Z - z)} \frac{(e^{-2\tau_c} - 1)(\alpha_{\rm out} + \sigma_{\rm out}/2 - \alpha_{\rm out}\mu)}{1 - (2/\sigma_{\rm out})(\alpha_{\rm out} + \sigma_{\rm out}/2 - \alpha_{\rm out}\mu)e^{-2\tau_c}} + 1\right].
\end{eqnarray}

Now I can compare the CR flux $F = \int I d\Omega = I_+ + I_-$ inside the cloud to the CR flux far away from the cloud.  At the center of the cloud, $z = 0$, the CR intensity is the intensity entering into the cloud after being attenuated: $I_+(0) = I_-(0) = I_+(Z) e^{\tau_c} = I_-(Z) e^{-\tau_c}$.  Far from the cloud, as $z \to \infty$, $I_+(\infty) = I_-(\infty) = Q_{\rm out} / (2\alpha_{\rm out})$, which simply means that the CR density is equal to their injection rate times their lifetime.  Taking the ratio of the fluxes, I have
\begin{equation}
\frac{F(0)}{F(\infty)} = \exp(-\tau_c) \frac{\mu - 1}{\sigma_{\rm out}/(2\alpha_{\rm out}) - (1 + \sigma_{\rm out}/(2\alpha_{\rm out}) - \mu)e^{-2\tau_c}}.
\end{equation}
In a galaxy, the CR scattering mean free path is thought to be much smaller than the absorbing mean free path, so that $\sigma_{\rm out} \gg \alpha_{\rm out}$.  This ratio is roughly
\begin{equation}
\label{eqn:FluxRatioSAtmos}
\frac{F(0)}{F(\infty)} \approx 2 \sqrt{\frac{\alpha_{\rm out}}{\sigma_{\rm out}}} \frac{e^{-\tau_c}}{1 - e^{-2\tau_c} + 2 \sqrt{\alpha_{\rm out}/\sigma_{\rm out}} e^{-2\tau_c}}.
\end{equation} 

As it turns out, even clouds that are optically thin to CR absorption on one pass can suppress the CR flux within them.  If $\tau_c \ll 1$, I have 
\begin{equation}
\frac{F(0)}{F(\infty)} \approx \sqrt{\frac{\alpha_{\rm out}}{\sigma_{\rm out}}} \frac{1 - \tau_c}{\sqrt{\alpha_{\rm out}/\sigma_{\rm out}} + (1 - 2\sqrt{\alpha_{\rm out}/\sigma_{\rm out}})\tau_c}.
\end{equation} 
For very small cloud absorption optical depths, the flux within the cloud is the same at infinity.  However, once $\tau_c$ reaches $\sim \sqrt{\alpha_{\rm out}/\sigma_{\rm out}} \ll 1$, the flux begins to drop inversely with $\tau_c$.  This is despite the fact that CRs free stream within the cloud in this model and are not trapped inside it: the flux is low because they can make multiple passes through a cloud and have multiple chances to be destroyed.

Equation~\ref{eqn:FluxRatioSAtmos} lets me solve for the approximate column (in the 1-D planar case) when gamma-ray ionization dominates over cosmic ray ionization.  Setting $F(0)/F(\infty) = \zeta_{\gamma}/\zeta_{\rm CR}^{\rm ext}$ where $\zeta_{\rm CR}^{\rm ext}$ is the CR ionization rate outside of the cloud, I get
\begin{equation}
\label{eqn:tauGRDR}
\tau_{\rm GRDR} = -\ln \left[\frac{-(\zeta_{\rm CR}^{\rm ext} / \zeta_{\gamma}) \sqrt{\alpha_{\rm out}/\sigma_{\rm out}} + \sqrt{\alpha_{\rm out}/\sigma_{\rm out} (\zeta_{\rm CR}^{\rm ext} / \zeta_{\gamma})^2 + 1 - 2\sqrt{\alpha_{\rm out}/\sigma_{\rm out}}}}{1 - 2\sqrt{\alpha_{\rm out}/\sigma_{\rm out}}}\right].
\end{equation}
This calculation neglects absorption of gamma rays within the cloud, which will start to matter over columns of $\sim 200\ \gcm2$ ($\tau_{\rm GRDR} \ga 1$).  In the limit that scattering outside the cloud is slow and the exterior CR ionization rate is much greater than the gamma-ray ionization rate, with $\sqrt{\alpha_{\rm out} / \sigma_{\rm out}} \ll \zeta_{\gamma} / \zeta_{\rm CR}^{\rm ext}$, equation~\ref{eqn:tauGRDR} can be simplified to
\begin{equation}
\tau_{\rm GRDR} \approx -\ln \left[\frac{(\zeta_{\gamma} / \zeta_{\rm CR}^{\rm ext})}{2 \sqrt{\alpha_{\rm out} / \sigma_{\rm out}}}\right].
\end{equation}


\begin{thebibliography}{}
\bibitem[Abdo et al.(2010)]{Abdo10-Starburst} Abdo, A.~A., et al.\ 2010, \apjl, 709, L152 

\bibitem[Abergel et al.(2011)]{Abergel11} Planck Collaboration, Abergel, A., Ade, P.~A.~R., et al.\ 2011, \aap, 536, A25 

\bibitem[Acciari et al.(2009)]{Acciari09} Acciari, V.~A., et al.\ 2009, \nat, 462, 770 

\bibitem[Acero et al.(2009)]{Acero09} Acero, F., et al.\ 2009, Science, 326, 1080 

\bibitem[Anantharamaiah et al.(2000)]{Anantharamaiah00} Anantharamaiah, K.~R., Viallefond, F., Mohan, N.~R., Goss, W.~M., \& Zhao, J.~H.\ 2000, \apj, 537, 613 

\bibitem[Bastian et al.(2010)]{Bastian10} Bastian, N., Covey, K.~R., \& Meyer, M.~R.\ 2010, \araa, 48, 339 

\bibitem[Bayet et al.(2011)]{Bayet11} Bayet, E., Williams, D.~A., Hartquist, T.~W., \& Viti, S.\ 2011, \mnras, 414, 1583

\bibitem[Berestetskii et al.(1979)]{Berestetskii79} Berestetskii, V. B., Lifshitz, E. M., \& Pitaevskii, L. P. 1979, \emph{Quantum Electrodynamics}, 2nd ed. (Oxford: Butterworth-Heinemann)

\bibitem[Bergin \& Tafalla(2007)]{Bergin07} Bergin, E.~A., \& Tafalla, M.\ 2007, \araa, 45, 339 

\bibitem[Brown et al.(2004)]{Brown04} Brown, R.~L., Wild, W., \& Cunningham, C.\ 2004, Advances in Space Research, 34, 555 

\bibitem[Cameron(1962)]{Cameron62} Cameron, A.~G.~W.\ 1962, Icarus, 1, 13 

\bibitem[Caselli et al.(1998)]{Caselli98} Caselli, P., Walmsley, C.~M., Terzieva, R., \& Herbst, E.\ 1998, \apj, 499, 234 

\bibitem[Caselli et al.(2002)]{Caselli02} Caselli, P., Walmsley, C.~M., Zucconi, A., et al.\ 2002, \apj, 565, 344 

\bibitem[Cesarsky \& Volk(1978)]{Cesarsky78} Cesarsky, C.~J., \& Volk, H.~J.\ 1978, \aap, 70, 367 

\bibitem[Chevalier \& Clegg(1985)]{Chevalier85} Chevalier, R.~A., \& Clegg, A.~W.\ 1985, \nat, 317, 44 

\bibitem[Condon(1992)]{Condon92} Condon, J.~J.\ 1992, \araa, 30, 575 

\bibitem[Cravens \& Dalgarno(1978)]{Cravens78} Cravens, T.~E., \& Dalgarno, A.\ 1978, \apj, 219, 750 

\bibitem[Crocker et al.(2011)]{Crocker11-Wild} Crocker, R.~M., Jones, D.~I., Aharonian, F., Law, C.~J., Melia, F., Oka, T., \& Ott, J.\ 2011, \mnras, 413, 763 

\bibitem[Crutcher(1999)]{Crutcher99} Crutcher, R.~M.\ 1999, \apj, 520, 706 

\bibitem[Davies et al.(2007)]{Davies07} Davies, R.~I., M{\"u}ller S{\'a}nchez, F., Genzel, R., et al.\ 2007, \apj, 671, 1388 

\bibitem[Diehl et al.(2006)]{Diehl06} Diehl, R., et al.\ 2006, \nat, 439, 45 

\bibitem[Downes \& Solomon(1998)]{Downes98} Downes, D., \& Solomon, P.~M.\ 1998, \apj, 507, 615 

\bibitem[Downes \& Eckart(2007)]{Downes07} Downes, D., \& Eckart, A.\ 2007, \aap, 468, L57 

\bibitem[Fatuzzo et al.(2006)]{Fatuzzo06} Fatuzzo, M., Adams, F.~C., \& Melia, F.\ 2006, \apjl, 653, L49 

\bibitem[Figer et al.(1999)]{Figer99} Figer, D.~F., Kim, S.~S., Morris, M., et al.\ 1999, \apj, 525, 750 

\bibitem[Fujita et al.(2009)]{Fujita09} Fujita, Y., Ohira, Y., Tanaka, S.~J., \& Takahara, F.\ 2009, \apjl, 707, L179 

\bibitem[Fujita et al.(2010)]{Fujita10} Fujita, Y., Ohira, Y., \& Takahara, F.\ 2010, \apjl, 712, L153 

\bibitem[Gabici et al.(2007)]{Gabici07} Gabici, S., Aharonian, F.~A., \& Blasi, P.\ 2007, \apss, 309, 365 

\bibitem[Gabici et al.(2010)]{Gabici10} Gabici, S., Casanova, S., Aharonian, F.~A., \& Rowell, G.\ 2010, SF2A-2010: Proceedings of the Annual meeting of the French Society of Astronomy and Astrophysics, 313 

\bibitem[Gammie(1996)]{Gammie96} Gammie, C.~F.\ 1996, \apj, 457, 355 

\bibitem[Ginzburg \& Ptuskin(1976)]{Ginzburg76} Ginzburg, V.~L., \& Ptuskin, V.~S.\ 1976, Reviews of Modern Physics, 48, 161 

\bibitem[Goetz et al.(1990)]{Goetz90} Goetz, M., Downes, D., Greve, A., \& McKeith, C.~D.\ 1990, \aap, 240, 52

\bibitem[Heckman et al.(2001)]{Heckman01} Heckman, T.~M., Sembach, K.~R., Meurer, G.~R., et al.\ 2001, \apj, 558, 56 

\bibitem[Heckman(2003)]{Heckman03} Heckman, T. M. 2003, in Rev. Mex. AA Ser. Conf., 17, 47

\bibitem[Hezareh et al.(2008)]{Hezareh08} Hezareh, T., Houde, M., McCoey, C., Vastel, C., \& Peng, R.\ 2008, \apj, 684, 1221 

\bibitem[Hollenbach \& Tielens(1999)]{Hollenbach99} Hollenbach, D.~J., \& Tielens, A.~G.~G.~M.\ 1999, Reviews of Modern Physics, 71, 173 

\bibitem[Hopkins et al.(2010)]{Hopkins10} Hopkins, P.~F., Murray, N., Quataert, E., \& Thompson, T.~A.\ 2010, \mnras, 401, L19 

\bibitem[Hopkins et al.(2012)]{Hopkins12} Hopkins, P.~F., Hayward, C.~C., Narayanan, D., \& Hernquist, L.\ 2012, \mnras, 420, 320 

\bibitem[Hurwitz et al.(1997)]{Hurwitz97} Hurwitz, M., Jelinsky, P., \& Dixon, W.~V.~D.\ 1997, \apjl, 481, L31 

\bibitem[Indriolo et al.(2009)]{Indriolo09} Indriolo, N., Fields, B.~D., \& McCall, B.~J.\ 2009, \apj, 694, 257 

\bibitem[Inoue(2011)]{Inoue11} Inoue, Y.\ 2011, \apj, 728, 11

\bibitem[Jones et al.(2008)]{Jones08} Jones, D.~I., Protheroe, R.~J., \& Crocker, R.~M.\ 2008, PASA, 25, 161 

\bibitem[Jones et al.(2011)]{Jones11} Jones, D.~I., Crocker, R.~M., Ott, J., Protheroe, R.~J., \& Ekers, R.~D.\ 2011, \aj, 141, 82 

\bibitem[Kamae et al.(2006)]{Kamae06} Kamae, T., Karlsson, N., Mizuno, T., Abe, T., \& Koi, T.\ 2006, \apj, 647, 692 

\bibitem[Kennicutt(1998)]{Kennicutt98} Kennicutt, R. C. 1998, \apj, 498, 541

\bibitem[Kim et al.(2006)]{Kim06} Kim, S.~S., Figer, D.~F., Kudritzki, R.~P., \& Najarro, F.\ 2006, \apjl, 653, L113 

\bibitem[Klessen et al.(2000)]{Klessen00} Klessen, R.~S., Heitsch, F., \& Mac Low, M.-M.\ 2000, \apj, 535, 887 

\bibitem[Krumholz et al.(2012)]{Krumholz12} Krumholz, M.~R., Dekel, A., \& McKee, C.~F.\ 2012, \apj, 745, 69 

\bibitem[Kulsrud \& Cesarsky(1971)]{Kulsrud71} Kulsrud, R.~M., \& Cesarsky, C.~J.\ 1971, \aplett, 8, 189 

\bibitem[Lacki et al.(2010)]{Lacki10-FRC1} Lacki, B.~C., Thompson, T.~A., \& Quataert, E.\ 2010, \apj, 717, 1 

\bibitem[Lacki \& Thompson(2010)]{Lacki10-XRay} Lacki, B.~C., \& Thompson, T.~A.\ 2010, arXiv:1010.3030 

\bibitem[Lacki et al.(2011)]{Lacki11-Obs} Lacki, B.~C., Thompson, T.~A., Quataert, E., Loeb, A., \& Waxman, E.\ 2011, \apj, 734, 107 

\bibitem[Lacki(2012)]{Lacki12-Al26} Lacki, B.~C., 2012, ApJ, submitted

\bibitem[Lehmer et al.(2010)]{Lehmer10} Lehmer, B.~D., Alexander, D.~M., Bauer, F.~E., Brandt, W.~N., Goulding, A.~D., Jenkins, L.~P., Ptak, A., \& Roberts, T.~P.\ 2010, \apj, 724, 559 

\bibitem[Larson(2005)]{Larson05} Larson, R.~B.\ 2005, \mnras, 359, 211 

\bibitem[Leitherer et al.(1995)]{Leitherer95} Leitherer, C., Ferguson, H.~C., Heckman, T.~M., \& Lowenthal, J.~D.\ 1995, \apjl, 454, L19 

\bibitem[Loeb \& Waxman(2006)]{Loeb06} Loeb, A. \& Waxman, E. 2006, Journal of Cosmology and Astroparticle Physics 5, 3

\bibitem[Maloney et al.(1996)]{Maloney96} Maloney, P.~R., Hollenbach, D.~J., \& Tielens, A.~G.~G.~M.\ 1996, \apj, 466, 561 

\bibitem[Mannheim \& Schlickeiser(1994)]{Mannheim94} Mannheim, K., \& Schlickeiser, R.\ 1994, \aap, 286, 983 

\bibitem[Mapelli et al.(2006)]{Mapelli06} Mapelli, M., Ferrara, A., \& Pierpaoli, E.\ 2006, \mnras, 369, 1719 

\bibitem[Maret \& Bergin(2007)]{Maret07} Maret, S., \& Bergin, E.~A.\ 2007, \apj, 664, 956 

\bibitem[McCrady et al.(2003)]{McCrady03} McCrady, N., Gilbert, A.~M., \& Graham, J.~R.\ 2003, \apj, 596, 240 

\bibitem[McKee(1989)]{McKee89} McKee, C.~F.\ 1989, \apj, 345, 782 

\bibitem[Meijerink \& Spaans(2005)]{Meijerink05} Meijerink, R., \& Spaans, M.\ 2005, \aap, 436, 397 

\bibitem[Meijerink et al.(2006)]{Meijerink06} Meijerink, R., Spaans, M., \& Israel, F.~P.\ 2006, \apjl, 650, L103 

\bibitem[Meijerink et al.(2011)]{Meijerink11} Meijerink, R., Spaans, M., Loenen, A.~F., \& van der Werf, P.~P.\ 2011, \aap, 525, A119 

\bibitem[Mestel \& Spitzer(1956)]{Mestel56} Mestel, L., \& Spitzer, L., Jr.\ 1956, \mnras, 116, 503 

\bibitem[Morfill(1982)]{Morfill82} Morfill, G.~E.\ 1982, \apj, 262, 749 

\bibitem[Mouschovias \& Spitzer(1976)]{Mouschovias76} Mouschovias, T.~C., \& Spitzer, L., Jr.\ 1976, \apj, 210, 326 

\bibitem[Murphy(2011)]{Murphy11} Murphy, E.~J.\ 2011, UP2010: Have Observations Revealed a Variable Upper End of the Initial Mass Function?, 440, 361 

\bibitem[Myers(1983)]{Myers83} Myers, P.~C.\ 1983, \apj, 270, 105 

\bibitem[Norris(1988)]{Norris88} Norris, R.~P.\ 1988, \mnras, 230, 345 

\bibitem[Ostriker et al.(2001)]{Ostriker01} Ostriker, E.~C., Stone, J.~M., \& Gammie, C.~F.\ 2001, \apj, 546, 980 

\bibitem[Padmanabhan \& Finkbeiner(2005)]{Padmanabhan05} Padmanabhan, N., \& Finkbeiner, D.~P.\ 2005, \prd, 72, 023508 

\bibitem[Papadopoulos(2010)]{Papadopoulos10-CRDRs} Papadopoulos, P.~P.\ 2010, \apj, 720, 226 

\bibitem[Papadopoulos et al.(2010)]{Papadopoulos10-A220} Papadopoulos, P.~P., Isaak, K., \& van der Werf, P.\ 2010, \apj, 711, 757 

\bibitem[Papadopoulos et al.(2011)]{Papadopoulos11-SF} Papadopoulos, P.~P., Thi, W.-F., Miniati, F., \& Viti, S.\ 2011, \mnras, 414, 1705 

\bibitem[Paumard et al.(2006)]{Paumard06} Paumard, T., Genzel, R., Martins, F., et al.\ 2006, \apj, 643, 1011 

\bibitem[Persic et al.(2004)]{Persic04} Persic, M., Rephaeli, Y., Braito, V., Cappi, M., Della Ceca, R., Franceschini, A., \& Gruber, D.~E.\ 2004, \aap, 419, 849 

\bibitem[Pohl(1994)]{Pohl94} Pohl, M.\ 1994, \aap, 287, 453 

\bibitem[Protheroe et al.(2008)]{Protheroe08} Protheroe, R.~J., Ott, J., Ekers, R.~D., Jones, D.~I., \& Crocker, R.~M.\ 2008, \mnras, 390, 683 

\bibitem[Ranalli et al.(2003)]{Ranalli03} Ranalli, P., Comastri, A., \& Setti, G.\ 2003, \aap, 399, 39 

\bibitem[Rengarajan(2005)]{Rengarajan05} Rengarajan, T.~N.\ 2005, Proc. 29th Int. Cosmic Ray Conf. (Pune), 3

\bibitem[Rieke et al.(1980)]{Rieke80} Rieke, G.~H., Lebofsky, M.~J., Thompson, R.~I., Low, F.~J., \& Tokunaga, A.~T.\ 1980, \apj, 238, 24 

\bibitem[Robishaw et al.(2008)]{Robishaw08} Robishaw, T., Quataert, E., \& Heiles, C.\ 2008, \apj, 680, 981 

\bibitem[Rybicki \& Lightman(1979)]{Rybicki79} Rybicki, G. B. \& Lightman, A. P. 1979, \emph{Radiative Processes in Astrophysics}, (New York: Wiley-VCH).

\bibitem[Sakamoto et al.(2008)]{Sakamoto08} Sakamoto, K., Wang, J., Wiedner, M.~C., et al.\ 2008, \apj, 684, 957 

\bibitem[Sanders et al.(2003)]{Sanders03} Sanders, D.~B., Mazzarella, J.~M., Kim, D.-C., Surace, J.~A., \& Soifer, B.~T.\ 2003, \aj, 126, 1607 

\bibitem[Schlickeiser(2002)]{Schlickeiser02} Schlickeiser, R. 2002, \emph{Cosmic Ray Astrophysics}, (New York: Springer)

\bibitem[Skilling \& Strong(1976)]{Skilling76} Skilling, J., \& Strong, A.~W.\ 1976, \aap, 53, 253 

\bibitem[Smith \& Gallagher(2001)]{Smith01} Smith, L.~J., \& Gallagher, J.~S.\ 2001, \mnras, 326, 1027 

\bibitem[Socrates et al.(2008)]{Socrates08} Socrates, A., Davis, S.~W., \& Ramirez-Ruiz, E.\ 2008, \apj, 687, 202 

\bibitem[Stecker(1970)]{Stecker70} Stecker, F.~W.\ 1970, \apss, 6, 377

\bibitem[Stepinski(1992)]{Stepinski92} Stepinski, T.~F.\ 1992, Icarus, 97, 130 

\bibitem[Stone et al.(1998)]{Stone98} Stone, J.~M., Ostriker, E.~C., \& Gammie, C.~F.\ 1998, \apjl, 508, L99 

\bibitem[Strong \& Moskalenko(1998)]{Strong98} Strong, A.~W., \& Moskalenko, I.~V.\ 1998, \apj, 509, 212 

\bibitem[Strong et al.(2010)]{Strong10} Strong, A.~W., Porter, T.~A., Digel, S.~W., J{\'o}hannesson, G., Martin, P., Moskalenko, I.~V., Murphy, E.~J., \& Orlando, E.\ 2010, \apjl, 722, L58 

\bibitem[Suchkov et al.(1993)]{Suchkov93} Suchkov, A., Allen, R.~J., \& Heckman, T.~M.\ 1993, \apj, 413, 542 

\bibitem[Tacconi et al.(2008)]{Tacconi08} Tacconi, L.~J., Genzel, R., Smail, I., et al.\ 2008, \apj, 680, 246 

\bibitem[Thompson, Quataert, \& Waxman(2007)]{Thompson07} Thompson, T. A., Quataert, E., Waxman, E. 2007, \apj, 654, 219

\bibitem[Torres(2004)]{Torres04} Torres, D. F. 2004, \apj, 617, 966

\bibitem[Torres et al.(2010)]{Torres10} Torres, D.~F., Marrero, A.~Y.~R., \& de Cea Del Pozo, E.\ 2010, \mnras, 408, 1257 

\bibitem[Umebayashi \& Nakano(1981)]{Umebayashi81} Umebayashi, T., \& Nakano, T.\ 1981, \pasj, 33, 617 

\bibitem[Umebayashi \& Nakano(2009)]{Umebayashi09} Umebayashi, T., \& Nakano, T.\ 2009, \apj, 690, 69 

\bibitem[van Dokkum \& Conroy(2010)]{vanDokkum10} van Dokkum, P.~G., \& Conroy, C.\ 2010, \nat, 468, 940 

\bibitem[van Dokkum \& Conroy(2011)]{vanDokkum11} van Dokkum, P.~G., \& Conroy, C.\ 2011, \apjl, 735, L13 

\bibitem[Williams et al.(1998)]{Williams98} Williams, J.~P., Bergin, E.~A., Caselli, P., Myers, P.~C., \& Plume, R.\ 1998, \apj, 503, 689 

\bibitem[Williams \& Bower(2010)]{Williams10} Williams, P.~K.~G., \& Bower, G.~C.\ 2010, \apj, 710, 1462 

\end{thebibliography}
\end{document}